\documentclass[10pt,twocolumn,journal]{IEEEtran}
\usepackage{latexsym}
\usepackage{float}
\usepackage{amsfonts}
\usepackage{amsbsy}
\usepackage{amssymb}
\usepackage{times}
\usepackage{graphicx}
\usepackage{setspace}
\usepackage{enumerate}
\usepackage[usenames]{color}
\usepackage[dvips]{pstcol}
\usepackage{epstopdf}
\usepackage{cite}
\usepackage{amssymb}
\usepackage{amsfonts}
\usepackage{graphicx}
\usepackage{epsfig}
\usepackage{psfrag}
\usepackage{xcolor}
\usepackage{amsfonts, bm}
\usepackage{epstopdf}
\usepackage{cite}
\usepackage{color}
\usepackage{xcolor}
\usepackage{verbatim}
\usepackage{multirow}
\usepackage{array}
\usepackage{booktabs}
\usepackage{amsthm}
\usepackage{makecell}
\usepackage{units}
\usepackage[linesnumbered, ruled]{algorithm2e}
\usepackage{algpseudocode}
\usepackage[ruled]{algorithm2e}
\usepackage{amsmath}
\usepackage{xurl}
\usepackage[colorlinks=true, linkcolor=black, citecolor=black, urlcolor=black]{hyperref}
\usepackage{subcaption}
\IEEEoverridecommandlockouts
\begin{document}
\title{F$^4$-CKM: Learning Channel Knowledge Map with Radio Frequency Radiance Field Rendering}
\author{\IEEEauthorblockN{Kequan Zhou, Guangyi Zhang, Hanlei Li, Yunlong Cai, Shengli Liu, and Guanding Yu}
	\thanks{ K. Zhou, G. Zhang, H. Li, Y. Cai, and G. Yu are with the College of Information Science and Electronic Engineering, Zhejiang University, Hangzhou 310027, China (e-mail: kqzhou@zju.edu.cn; zhangguangyi@zju.edu.cn; hanleili@zju.edu.cn; ylcai@zju.edu.cn; yuguanding@zju.edu.cn).
		
	S. Liu is with the School of Communication and Information Engineering, Shanghai University, Shanghai, China. (email: victoryliu@shu.edu.cn).}}
%

\maketitle
\vspace{-3.3em}
\begin{abstract}
In 6G mobile communications, acquiring accurate and timely channel state information (CSI) becomes increasingly challenging due to the growing antenna array size and bandwidth.
To alleviate the CSI feedback burden, the channel knowledge map (CKM) has emerged as a promising approach by leveraging environment-aware techniques to predict CSI based solely on user locations.
However, how to effectively construct a CKM remains an open issue.
In this paper, we propose F$^4$-CKM, a novel CKM construction framework characterized by four distinctive features: radiance Field rendering, spatial-Frequency-awareness, location-Free usage, and Fast learning.
Central to our design is the adaptation of radiance field rendering techniques from computer vision to the radio frequency (RF) domain, enabled by a novel Wireless Radiator Representation (WiRARE) network that captures the spatial-frequency characteristics of wireless channels.
Additionally, a novel shaping filter module and an angular sampling strategy are introduced to facilitate CKM construction.
Extensive experiments demonstrate that F$^4$-CKM significantly outperforms existing baselines in terms of wireless channel prediction accuracy and efficiency.
\end{abstract}

\IEEEpeerreviewmaketitle

\section{Introduction}\label{intro}
Accurate and timely channel state information (CSI) is essential for ensuring reliable and high-throughput wireless communications.
Traditional CSI acquisition methods involve transmitting pilot signals and obtaining estimated CSI via feedback links \cite{Wu2024transformer,Zhang2024scan,Chen2023viewing}.
However, with 6G communications expected to deploy large-scale multiple-input multiple-output (MIMO) systems and operate over wide bandwidths \cite{Wang2021a, Lu2022communicating, You2025next, Wang2024dynamic, Zhao2025energy, Zhou2025feature, Wang2025a}, the resulting overhead from conventional approaches becomes prohibitive.
To address this, the channel knowledge map (CKM) has emerged as a promising alternative \cite{Zeng2021toward, Zeng2024a, Liu2025channel}.
By leveraging environment-aware techniques, CKM enables CSI prediction directly from user locations, significantly reducing CSI acquisition overhead.
Despite its potential to enhance future 6G networks, the effective construction of CKM remains an open and critical challenge requiring further investigation.

\subsection{Prior Work}
CKM construction methods can generally be categorized into two main classes: model-based and model-free approaches \cite{Xie2024on}, distinguished primarily by whether they incorporate expert knowledge of wireless channel models.
Model-based approaches leverage domain knowledge, making them inherently explainable, configurable, and less reliant on large-scale training data \cite{Karttunen2017spatially,Kanhere2023calibration,Charbonnier2020calibration,Li2022channel}.
Specifically, the authors in \cite{Kanhere2023calibration} proposed a ray tracing model to simulate wireless signal propagation, enabling material parameter calibration for enhanced accuracy.
Moreover, the authors in \cite{Li2022channel} utilized the expectation-maximization algorithm to derive closed-form expressions with only a few tunable parameters, resulting in efficient CKM construction.
Despite these advantages, model-based approaches often face performance limitations due to discrepancies between theoretical models and complex real-world environments.
On the other hand, model-free methods do not depend on expert priors but instead exploit data collection and deep learning techniques to improve performance \cite{Liu2021fire,Fu2024generative, Si2025unsupervised,Hehn2024differentiable,Jin2025an,Sifaou2025semi}.
In particular, a generative approach based on the Laplacian pyramid was proposed in \cite{Jin2025an}, which hierarchically constructs a CKM by extracting multi-frequency components from a morphological map. 
Additionally, \cite{Sifaou2025semi} proposed a cross-prediction-powered, inference-based method that effectively utilizes both labeled and unlabeled data during training.
Although model-free methods have demonstrated promising performance, they are often hindered by data scarcity, as acquiring high-quality field measurements is both time-consuming and expensive.

To leverage the advantages of both model-based and model-free methods, hybrid methods have gained increasing attention \cite{Orekondy2023winert,Jin2024sandwich,Hoydis2024learning,Chen2024diffraction,Xiao2024from}.
These methods integrate domain knowledge with data-driven learning to enhance both accuracy and interpretability.
For instance, \cite{Orekondy2023winert} explored neural ray tracing, where neural networks (NNs) are employed to iteratively learn ray-scene interactions.
Meanwhile, the authors in \cite{Hoydis2024learning} proposed Neural Materials, a neural ray tracing framework in which all trainable parameters have precise physical meanings.
Beyond neural ray tracing, a virtual obstacle model was developed in \cite{Chen2024diffraction}.
This model effectively captures the diffraction and scattering mechanisms with the aid of NNs, thereby learning an explainable CKM.
Furthermore, the authors in \cite{Xiao2024from} formulated CKM construction as an ordinary differential equation (ODE) problem, exploiting the spatial correlation of wireless channels.
The ODE function is implicitly represented by a physics-inspired NN, which learns the gradients necessary for solving the ODE.

Recent advances in computer graphics have further motivated the research of hybrid CKM construction methods \cite{Wu2024embracing}.
Neural radiance field (NeRF) \cite{Mildenhall2021nerf} and three-dimensional Gaussian splatting (3DGS) \cite{Kerbl20233d} have emerged as powerful techniques for 3D scene representation through visual radiance field rendering.
These approaches model 3D environments using parameterized spatial units and treat multi-view images as training data, enabling the synthesis of novel views by learning the light distribution within a scene.
Inspired by the fact that both wireless signals and visible light are electromagnetic (EM) waves, recent works have explored the integration of radiance field rendering techniques with wireless channel modeling to construct 3D CKMs \cite{Zhao2023nerf2,Zhang2024rf,Lu2024newrf,Wen2024wrf}.
Specifically, a pioneering NeRF-inspired scheme known as NeRF$^2$ was introduced in \cite{Zhao2023nerf2} for CKM construction.
This scheme takes into account the unique characteristics of radio frequency (RF) signals and adapts the NeRF framework for wireless channel modeling, thus enhancing CKM construction accuracy.
In \cite{Lu2024newrf}, the authors proposed NeWRF, which enhances data efficiency through a coarse-to-fine training strategy and leverages estimated angle-of-arrival (AoA) information as an auxiliary input, enabling CKM construction even with limited data. 
Furthermore, \cite{Wen2024wrf} introduced a 3DGS-based method named WRF-GS+, which leverages spherical and Mercator projections to adapt the optical camera model to an RF antenna model, thereby enabling effective CKM construction.

\subsection{Motivation and Contributions}\label{Motivation}
Despite recent advancements, existing techniques for CKM construction face three key limitations:

\begin{itemize}
	\item \textit{Lack of spatial-frequency awareness:}
	Existing RF radiance field modeling and NN designs have not fully considered the characteristics of wireless channels.
	In the spatial domain, antennas in close proximity exhibit strong spatial correlations; similarly, subcarriers in the frequency domain are inherently correlated \cite{Chen2023viewing}.
	These correlations are not adequately captured in existing methods, resulting in compromised performance.
	\item \textit{Lack of practical input:}
	Most existing approaches rely on accurate user locations as input for CKM-based CSI prediction.
	Nevertheless, it is difficult to obtain precise location in practice.
	For example, civilian global positioning system (GPS) systems typically offer meter-level accuracy \cite{Lee2016comparison}, which is quite coarse relative to the sub-wavelength resolution required by sub-6 GHz and millimeter-wave systems.
	As a result, the inaccurate location input may severely degrade prediction performance.
	Moreover, the use of location information can also draw privacy concerns and require encrypted transmission, bringing in additional overhead and potential privacy issues \cite{Yang2023environment}.
	\item \textit{Lack of construction efficiency:}
	Location-based methods often suffer from low construction efficiency, particularly in high-dimensional settings such as massive MIMO and orthogonal frequency division multiplexing (OFDM) systems.
	This inefficiency stems from the low-dimensional nature of location inputs, which provide limited information for learning.
	While positional encoding techniques can partially mitigate this issue, they also make the model more susceptible to positioning errors, thereby increasing the challenges of practical deployment.
	
\end{itemize}

In this paper, we address the limitations of prior CKM construction methods by introducing F$^4$-CKM.
To tackle the first limitation, we propose a novel RF radiance field model that effectively captures the spatial-frequency characteristics of wireless channels.
This model represents the 3D signal propagation environment via numerous spatial units, referred to as virtual wireless radiators, which are parameterized with spatial-frequency awareness. 
Subsequently, to render this RF radiance field, we develop the \textbf{Wi}reless \textbf{Ra}diator \textbf{Re}presentation (WiRARE) network to learn the parameters of these radiators.
The network backbone is specifically crafted to exploit spatial correlations in the channel, and is further enhanced with a series of frequency context awareness (FCA) modules that leverage frequency-domain dependencies.

To address the second limitation, we use uplink CSI as the input to CKM instead of relying on user location.
Prior studies \cite{Vieira2017deep,Xie2019md} have shown that a strong bidirectional mapping exists between uplink CSI and user location, owing to rich multipath propagation and large-scale antenna arrays.
Therefore, it is reasonable for a CKM to perform prediction based on uplink CSI.
Moreover, uplink CSI estimation techniques have already been well-established in the industry and do not pose privacy concerns, making it a practical and secure alternative.
Specifically, we propose a novel {shaping filter} module to perform data augmentation, facilitating the training of the WiRARE network on uplink CSI inputs.
This module boosts up CKM model convergence, effectively addressing the third limitation.
To further enhance construction efficiency, we introduce an improved angular sampling strategy, which leverages the spherical Fibonacci grid (SFG) \cite{Swinbank2006fibonacci} technique to achieve more effective angular sampling.  

In summary, the key contributions of this paper are as follows:

\begin{itemize}
\item We propose F$^4$-CKM, a novel CKM construction framework with four distinct features: radiance \textbf{F}ield rendering, spatial-\textbf{F}requency awareness, location-\textbf{F}ree usage, and \textbf{F}ast learning.
This unified design effectively addresses the major limitations of existing approaches, improving both the accuracy and efficiency of CKM construction.
\item We devise a radiance field rendering-based CKM construction method by formulating a novel RF radiance field model that captures the spatial-frequency characteristics of wireless channels. 
To render this field, we develop the WiRARE network, featuring a spatially aware backbone design and enhanced frequency sensitivity through embedded FCA modules.
\item Rather than the accurate user location, we propose to use uplink CSI as the CKM inputs, which will not be affected by positioning errors.
A novel shaping filter module is designed to perform data augmentation on uplink CSI inputs, enabling location-free CKM usage and accelerated model convergence.
Additionally, we adopt a more efficient angular sampling strategy based on the SFG to further enhance construction efficiency.
\item We conduct extensive experiments under diverse conditions across both simulated and real-world datasets to validate the effectiveness of our methods.
The results demonstrate that our F$^4$-CKM yields significant performance gains compared with existing baselines in terms of CKM construction accuracy and efficiency.
\end{itemize}

\subsection{Organization and Notation}
The remainder of the paper is organized as follows.
Section \ref{preliminaries} introduces the fundamentals of wireless channel modeling and NeRF techniques.
Section \ref{systemModel} describes the system model underlying the F$^4$-CKM framework.
The detailed designs of F$^4$-CKM are presented in Section \ref{methods}.
Then, simulation results are provided in Section \ref{Simulation}, and Section \ref{Conclusion} concludes the paper.

Unless otherwise specified, scalars, vectors, and tensors of dimension greater than one are denoted by lowercase, boldface lowercase, and boldface uppercase letters, respectively.
For a vector $\mathbf{a}$, $\mathbf{a}^H$ is its conjugate transpose.
For vectors and tensors, $||\cdot||$ denotes their Euclidean norm.
Finally, $\mathbb{C}^{m\times n}(\mathbb{R}^{m\times n})$ is the space of $m\times n$ complex (real) matrices.

\section{Preliminaries}\label{preliminaries}
In this section, we provide an overview of the fundamentals of wireless channel modeling and NeRF techniques.
\subsection{Wireless Channel Modeling}\label{SecWirelessChannelModel}
A signal transmitted from a single antenna at carrier frequency $f$ can be represented as $x(f)=Ae^{j\phi}$, where $A$ denotes the amplitude and $\phi$ is the phase.
In systems with multiple antennas and subcarriers, the transmitted signal is expressed as $\mathbf{X} = [\mathbf{x}(f_1),...,\mathbf{x}(f_{N_c})]^T\in\mathbb{C}^{N_c\times N_b}$, where $N_c$ is the number of subcarriers and $N_b$ is the number of transmit antennas\footnote{Without loss of generality, we consider the downlink scenario where the base station acts as the transmitter with $N_b$ antennas, while the user equipment serves as the receiver with $N_u$ antennas. The uplink case can be treated analogously by reversing the roles of the transmitter and receiver.}.
During propagation, the signal experiences both free space loss and multipath effects.
In free space propagation, the signal's amplitude attenuates inversely with the propagation distance $d$, while its phase rotates proportionally to $d$.
These effects are mathematically expressed as
\begin{IEEEeqnarray}{rCl}
	\Delta A = \frac{c}{4\pi fd}, \Delta \phi = \frac{-2\pi fd}{c}, \label{freeSpaceLoss}
\end{IEEEeqnarray}
respectively, where $c$ is the speed of light.
Then, the received signal at frequency $f$ can be expressed as
\begin{IEEEeqnarray}{rCl}
	y(f)=x(f)\Delta Ae^{j\Delta \phi}=A\cdot\Delta Ae^{j(\phi+\Delta \phi)}.
\end{IEEEeqnarray}
Taking into account the multipath effect, where the signal arrives at the receiver via multiple paths, each path introduces its own attenuation and phase shift.
As a result, the received signal $y(f)$ can be better described as
\begin{IEEEeqnarray}{rCl}
	y(f)=x(f)\sum_{l=1}^{L}\Delta A^le^{j\Delta \phi^l},
\end{IEEEeqnarray}
where $L$ is the number of propagation paths, $\Delta A^l$ and $e^{j\Delta \phi^l}$ represent the attenuation and phase shift of the $l$-th path, respectively.
The wireless channel model is defined as the cumulative effect of the environment on the transmitted signal and is given by
\begin{IEEEeqnarray}{rCl}
	h(f) \triangleq \frac{y(f)}{x(f)} = \sum_{l=1}^{L}\Delta A^le^{j\Delta \phi^l}.
\end{IEEEeqnarray}

In multiple-antenna systems, signals transmitted through different transmit-and-receive antenna pairs generally propagate through similar physical paths.
However, due to the spatial separation between antennas, the propagation distances of these paths can differ at the scale of the carrier wavelength.
These distance variations lead to path-dependent differences in signal attenuation and phase rotation. 
Therefore, the received signal $\mathbf{Y}\in\mathbb{C}^{N_c\times N_u}$ can be expressed as
\begin{IEEEeqnarray}{rCl}
	\mathbf{Y} & = & [\mathbf{y}(f_1),...,\mathbf{y}(f_{N_c})]^T, \IEEEyesnumber\IEEEyessubnumber*\\
	\mathbf{y}(f) & = & (\sum_{l=1}^{L}\Delta \mathbf{A}^l(f)\odot\exp(j\Delta \bm{\Phi}^l(f)))\mathbf{x}(f),
\end{IEEEeqnarray}
where $\Delta \mathbf{A}^l(f)\in\mathbb{R}^{N_u\times N_b}$ and $\exp(j\Delta \bm{\Phi}^l(f))\in\mathbb{C}^{N_u\times N_b}$ represent the attenuation and phase shifts of the $l$-th path, respectively, and $\odot$ denotes the element-wise multiplication operator for matrices.
The corresponding frequency-domain MIMO channel matrix is then given by
\begin{IEEEeqnarray}{rCl}\label{WirelessChannelModel}
	\mathbf{H}(f) & \triangleq & \frac{1}{P} \mathbf{y}(f)\mathbf{x}^H(f)\IEEEnonumber\\
	& = & \sum_{l=1}^{L}\Delta \mathbf{A}^l(f)\odot\exp(j\Delta \bm{\Phi}^l(f)),
\end{IEEEeqnarray}
where $P=\mathbf{x}^H(f)\mathbf{x}(f)$ denotes the transmit power.
The full channel representation over all subcarriers is then
\begin{IEEEeqnarray}{rCl}
	\mathbf{H} & = & [\mathbf{H}(f_1),...,\mathbf{H}(f_{N_c})]^T.
\end{IEEEeqnarray}
\subsection{Neural Radiance Field (NeRF)}
NeRF is a deep learning-based framework for representing 3D scenes via radiance field rendering.
It models a 3D scene using a continuous function learned by an NN.
This function takes as input a 3D spatial coordinate $\mathbf{p}=(p_x,p_y,p_z)$ and a viewing direction $\mathbf{r}=(r_\theta,r_\phi)$, and outputs the corresponding directional volume radiance $\mathbf{c}=(r, g, b)$ and volume density $\sigma$.
Here, the density $\sigma$ indicates the transmittance and opacity of a volume block.
Specifically, transmittance refers to the ratio of light that passes through volume blocks without being absorbed, while opacity describes the proportion of light that is absorbed and then radiated by volume blocks.

To render images from novel viewpoints, NeRF traces rays from each image pixel and samples a set of 3D points along each ray.
These sampled coordinates are fed into the NN to predict the corresponding radiance $\mathbf{c}$ and volume density $\sigma$.
The density $\sigma$ is then transformed into opacity $\alpha$.
The rendered pixel color $\hat{C}(\mathbf{r})$ can be calculated via the conventional $\alpha$-blending approach, given as
\begin{IEEEeqnarray}{CCl}
	\hat{C}(\mathbf{r})=\sum_{i=1}^{N}T_i\alpha_i\mathbf{c}_i,\IEEEyesnumber\IEEEyessubnumber*\\
	T_i=\prod_{j=1}^{i-1}(1-\alpha_j),\quad\alpha_i=1-e^{-\sigma_i\delta_i}, \label{nerfTalpha}
\end{IEEEeqnarray}
where $i$ and $N$ indicate the sample index and the total number of sampled points along the ray $\mathbf{r}$, respectively, $T$ represents the accumulated transmittance, and $\delta_i=||\mathbf{p}_{i+1}-\mathbf{p}_i||$ denotes the distance between adjacent samples.
After computing the color for each pixel, the full image is synthesized.
The synthesized image is then compared against the ground-truth image to compute a reconstruction loss, which is used to optimize the NN parameters.

As shown in Eq. (\ref{WirelessChannelModel}), wireless channels are typically modeled as the aggregate distortion from multiple propagation paths, akin to rendering images from light contributions along various viewing directions.
This analogy, coupled with NeRF's remarkable performance in 3D scene representation, motivates our adoption of NeRF techniques for wireless channel modeling.
By integrating NeRF in a hybrid fashion, we aim to enable more accurate construction of CKMs.

\section{RF Radiance Field Modeling} \label{systemModel}
In this section, we outline the formulation of the RF radiance field model underlying the F$^4$-CKM framework.
The objective is to predict the downlink CSI at the user equipment (UE) by leveraging the uplink CSI acquired at the base station (BS).
We focus on a wireless MIMO-OFDM system, where the BS is equipped with $N_b$ antennas, the UE has $N_u$ antennas, and the system operates over $N_c$ OFDM subcarriers.
To enable accurate channel prediction, we aim to learn a radiance field that captures the RF signal distribution in the surrounding environment.
Specifically, the adaptations of NeRF techniques for transitioning from the optical domain to the RF domain are detailed in Section \ref{basicField}.
Following that, we propose the formulation of the MIMO-OFDM RF radiance field in Section \ref{proposedField}.

\subsection{RF Radiance Field} \label{basicField}
For a clear explanation, the basic RF radiance field for a single-antenna, single-carrier system is first analyzed.
To begin with, the environment should first be discretized into a set of volume blocks.
According to the Huygens-Fresnel principle \cite{Zhao2023nerf2}, when the original RF signal $x$ transmitted by the BS reaches a volume block through multiple propagation paths, this block can be treated as a secondary source that {re-radiates} $x$.
Based on this principle, each volume block is modeled as a virtual wireless radiator, which absorbs RF signals from all directions and {re-radiates} them with an attenuation.
This attenuation is characterized by the radiator's transmittance $\beta$ and absorption ratio $\alpha = 1-\beta$, both of which are determined by the radiator's material property $\sigma$ and block length $\delta$.
A larger $\sigma$ or $\delta$ corresponds to a higher absorption ratio $\alpha$.
Each wireless radiator is characterized by a continuous function $\mathcal{F}_{\bm{\Theta}}(\cdot)$ learned by an NN, where $\bm{\Theta}$ denotes the NN's parameter set.
This function takes the spatial coordinate $\mathbf{p}$ and radiance direction $\bm{\omega}$ of each wireless radiator as inputs, and outputs the corresponding radiated signal $s$ and material property $\sigma$, given as
\begin{IEEEeqnarray}{rCl}
	\mathcal{F}_{\bm{\Theta}}(\cdot): (\mathbf{p},\bm{\omega})\longrightarrow(s,\sigma).
\end{IEEEeqnarray}

\begin{figure}[t]
	\centering
	\includegraphics[width=0.49\textwidth]{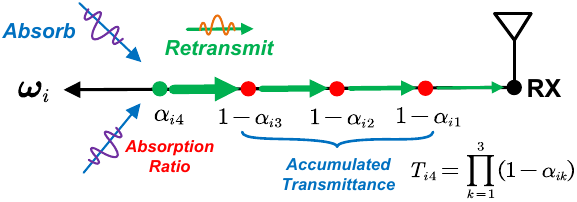}
	\captionsetup{font=footnotesize}
	\caption{Propagation modeling for the radiated signal from a wireless radiator.}
	\label{synthesize}
\end{figure}

To predict the downlink CSI, a set of rays $\{\bm{\omega}_i\}_{i=1}^{N_a}$ originating from the UE's antenna array should be sampled first, where $N_a$ denotes the number of rays across the angular domain.
Along each ray, a sequence of wireless radiators is sampled at intervals $\{\delta_{ij}\}_{j=1}^{N_r}$, where $i$ indicates the ray index and $N_r$ represents the number of sampled radiators along the radial direction.
With the ray direction and spatial coordinate of each sampled wireless radiator, the corresponding radiated signal and material property can be obtained from the NN $\mathcal{F}_{\bm{\Theta}}(\cdot)$.
Similar to Eq. (\ref{nerfTalpha}), the material property $\sigma$ is converted into the absorption ratio $\alpha$ and accumulated transmittance $T$, given as
\begin{IEEEeqnarray}{rCl}
	T_{ij}=\prod_{k=1}^{j-1}\beta_{ik},\quad\alpha_{ij}=1-e^{-\sigma_{ij}\delta_{ij}},\label{TandAlpha}
\end{IEEEeqnarray}
where $i$ denotes the ray index, both $j$ and $k$ indicate the radiator index along each ray, and $\beta=1-\alpha$ is the radiator's transmittance.
The propagation of the radiated signal from a wireless radiator is illustrated in Fig. \ref{synthesize}.
Subsequently, the received signal at the UE is computed as
\begin{IEEEeqnarray}{rCl}
	\hat{y} = \sum_{i=1}^{N_a}\sum_{j=1}^{N_r}T_{ij}\alpha_{ij}s_{ij}.
\end{IEEEeqnarray}
Finally, the downlink CSI can be expressed as
\begin{IEEEeqnarray}{rCl}
	\hat{h}^\mathrm{D} = \frac{\hat{y}}{x} = \sum_{i=1}^{N_a}\sum_{j=1}^{N_r}T_{ij}\alpha_{ij}c_{ij},\label{basicFieldChannel}
\end{IEEEeqnarray}
where $c_{ij} = s_{ij}/x$ represents the effect of aggregating all the RF signals arriving at the wireless radiator.
Since we are interested in the downlink CSI rather than the received signal, the radiated signal $s$ in the NN's outputs can be changed to the radiator aggregating coefficient $c$.
{As noted in Section~\ref{intro}, the use of explicit location information is susceptible to positioning errors and may raise privacy concerns.
To mitigate this issue, we replace the radiator coordinate $\mathbf{p}$ with a generic radiator-specific query $\mathbf{q}$, which can represent any available radiator-specific side information.
As a result, the NN's mapping function $\mathcal{F}_{\bm{\Theta}}(\cdot)$ can be represented as
\begin{IEEEeqnarray}{rCl}
	\mathcal{F}_{\bm{\Theta}}(\cdot): (\mathbf{q},\bm{\omega})\longrightarrow(c,\sigma),\label{basicFieldMapping}
\end{IEEEeqnarray}
where $\mathbf{q}$ abstracts radiator identity, facilitating robustness against positioning errors and enhancing privacy preservation.}

\subsection{MIMO-OFDM RF Radiance Field} \label{proposedField}
Building upon the basic formulation, we commence by delving into the formulation of MIMO-OFDM RF radiance fields.
In this context, two key challenges remain to be addressed:
\begin{itemize}
	\item [(i)] RF signals span multiple subcarrier frequencies, each of which experiences different free space propagation losses and interacts uniquely with the environment.
	Therefore, the formulation must account for these frequency-dependent effects to accurately model signal behavior.
	\item [(ii)] The transmitter and receiver are equipped with antenna arrays composed of multiple elements, necessitating the formulation to capture inter-element interactions, which substantially increases the problem’s complexity.
\end{itemize}

\subsubsection{Frequency-Aware Formulation}
{
To address the first challenge, we first redefine the formulations in Eq. (\ref{TandAlpha}) as
\begin{IEEEeqnarray}{rCl}
	T_{ij}(f)=\prod_{k=1}^{j-1}\beta_{ik}(f),\quad\alpha_{ij}(f) = 1-e^{-\sigma_{ij}(f)\delta_{ij}},\label{alpha_fd}
\end{IEEEeqnarray}
where $f$ denotes the subcarrier frequency and $\sigma_{ij}(f)$ aims to capture the frequency-dependent wireless radiator response.
However, Eq. (\ref{alpha_fd}) remains physically coupled, where the block length $\delta$ simultaneously governs two distinct effects:
\begin{itemize}
	\item [(i)] \textit{free-space propagation loss} (energy dispersion due to spherical wavefront spreading), which is independent of material properties, and
	\item [(ii)] \textit{within-medium attenuation} (material absorption over distance $\delta_{ij}$), which depends on $\sigma_{ij}(f)$.
\end{itemize}
\begin{figure}[t]
	\centering
	\includegraphics[width=0.49\textwidth]{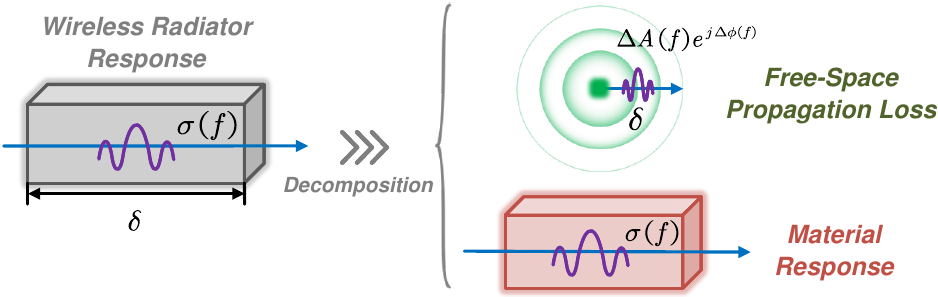}
	\captionsetup{font=footnotesize}
	\caption{Decomposition of the wireless radiator response.}
	\label{decomposition}
\end{figure}
Therefore, it is reasonable to embed domain knowledge of free-space propagation effects into the radiance field model, thereby improving generalization to new environments.
To this end, as shown in Fig. \ref{decomposition}, we decouple the free-space propagation effects from the per-radiator response.
Specifically, free-space amplitude attenuation and phase rotation are modeled explicitly as separate factors following Eq. (\ref{freeSpaceLoss}), given by
\begin{IEEEeqnarray}{lcl}
	\Delta A_{ij} = \frac{c}{4\pi f\sum_{k=1}^{j-1}\delta_{ik}}, \Delta \phi_{ij} = \frac{-2\pi f\sum_{k=1}^{j-1}\delta_{ik}}{c}. \label{freeSpaceLoss_new}
\end{IEEEeqnarray}
In addition, the absorption ratio $\alpha_{ij}(f)$ is redefined to reflect material-induced response alone.
Accordingly, $\sigma(f)$ is interpreted as an effective material response, with the explicit dependence on the geometric length $\delta$ eliminated, given by
\begin{IEEEeqnarray}{rCl}
	\alpha_{ij}(f)=1-e^{-\sigma_{ij}(f)}. \label{alpha_f}
\end{IEEEeqnarray}
Consequently, the accumulated attenuation factor $T_{ij}(f)$ must combine both the material-induced transmittance $\prod_{k=1}^{j-1}\Bigl(1-\alpha_{ik}(f)\Bigr)$ and the free-space propagation effects along the path, given by
\begin{IEEEeqnarray}{rCl}
	T_{ij}(f) & = & \Delta A_{ij}e^{j\Delta \phi_{ij}} \prod_{k=1}^{j-1}\Bigl(1-\alpha_{ik}(f)\Bigr) \IEEEyesnumber\IEEEyessubnumber* \\
	& = & \Delta A_{ij}e^{j\frac{-2\pi f\sum_{k=1}^{j-1}\delta_{ik}}{c}} \prod_{k=1}^{j-1}\beta_{ik}(f) \\
	& = & \Delta A_{ij}\prod_{k=1}^{j-1}\beta_{ik}(f)e^{j\frac{-2\pi f\delta_{ik}}{c}} \\
	& = & \frac{c}{4\pi f\sum_{k=1}^{j-1}\delta_{ik}}\prod_{k=1}^{j-1}\beta_{ik}(f)e^{j\frac{-2\pi f\delta_{ik}}{c}}. \label{T_f}
\end{IEEEeqnarray}
}
\subsubsection{Spatial-Aware Formulation}

\begin{figure}[t]
	\centering
	\includegraphics[width=0.49\textwidth]{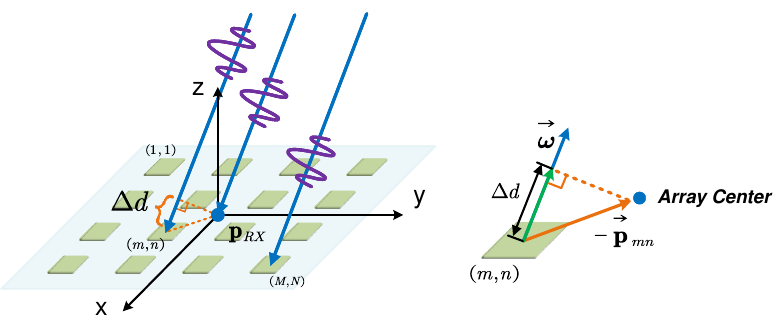}
	\captionsetup{font=footnotesize}
	\caption{Variations in the propagation distances of signals arriving at different receive antennas.}
	\label{spatialFormulation}
\end{figure}

Regarding the second challenge, we focus on a single subcarrier and denote the transmitted and received RF signals as $\mathbf{x}(f)\in\mathbb{C}^{N_b}$ and $\mathbf{y}(f)\in\mathbb{C}^{N_u}$, respectively.
To reveal the interactions between individual antenna elements, we model each radiator as a cluster of $N_u\times N_b$ sub-radiators, where each sub-radiator corresponds to a unique transmit-receive antenna pair.
Hence, the material property of the $j$-th radiator in the $i$-th ray is represented as a matrix $\bm{\sigma}_{ij}(f)\in\mathbb{R}^{N_u\times N_b}$, composed of the individual sub-radiators’ material properties.
{Similarly, the absorption ratio and transmittance are generalized as $\bm{\alpha}_{ij}(f) = \mathbf{J}- \exp\Bigl(\bm{\sigma}_{ij}(f)\Bigr)$ and $\bm{\beta}_{ij}(f) = \mathbf{J}- \bm{\alpha}_{ij}(f)$, respectively, where $\mathbf{J}\in\mathbb{R}^{N_u\times N_b}$ represents an all-ones matrix.}
Furthermore, the radiated signal is denoted as $\mathbf{s}_{ij}(f)\in\mathbb{C}^{N_u}$.
Then, the radiator aggregating coefficients can be derived as a matrix $\mathbf{C}_{ij}(f)\in\mathbb{C}^{N_u\times N_b}$, given by
\begin{IEEEeqnarray}{rCl}
	\mathbf{C}_{ij}(f) =\frac{1}{P} \mathbf{s}_{ij}(f)\mathbf{x}^H(f),
\end{IEEEeqnarray}
where $P=\mathbf{x}^H(f)\mathbf{x}(f)$ denotes the transmit power of the BS.
Since the antenna array {aperture} is typically much smaller than the signal propagation distance, signals arriving at different antennas can be treated as parallel.
Consequently, as illustrated in Fig. \ref{spatialFormulation}, the spatial separation between antennas causes slight differences in propagation distances, typically at the scale of the signal wavelength.
While these variations have negligible {effects} on signal amplitudes, they can induce significant phase shifts, resulting in large fluctuations in the predicted downlink CSI.
Therefore, it is crucial to calibrate the propagation distances for signals arriving at each individual receive antenna to mitigate such modeling errors.

\begin{figure*}[t]
	\centering
	\includegraphics[width=0.99\textwidth]{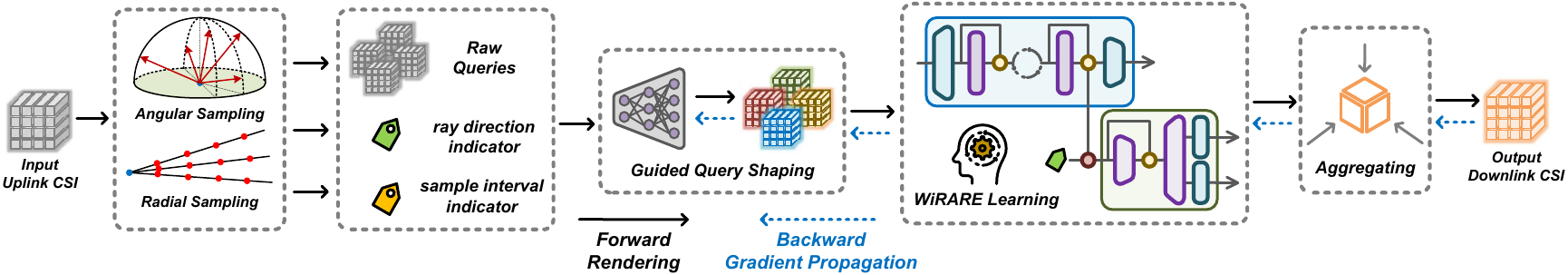}
	\captionsetup{font=footnotesize}
	\caption{The construction pipeline of F$^4$-CKM. The uplink CSI at the UE's antenna array is taken as input. We first perform angular sampling, followed by radial sampling along each ray. Subsequently, replicas of the original uplink CSI are fed into the {shaping filter} module, along with two flows of side information that serve as guidance. This module processes the CSI replicas into augmented queries for the WiRARE network. WiRARE then leverages these queries to obtain the aggregating coefficients and material properties of the sampled radiators, which are aggregated to form the final prediction for the downlink CSI.}
	\label{pipeline}
\end{figure*}

Specifically, we consider a UE equipped with a uniform planar array (UPA) consisting of $M$ rows and $N$ columns of antennas, where $M\times N=N_u$, and the antenna spacing is predetermined.
As shown in Fig. \ref{spatialFormulation}, our objective is to determine the propagation distance variation $\Delta d$ for the signal paths arriving at each receive antenna, with respect to the path towards the array center.
For the antenna $(m,n)$ located at the $m$-th row and the $n$-th column in the array, the distance calibration term is given by:
\begin{IEEEeqnarray}{rCl}
	\Delta d_{m,n} = \frac{-\overrightarrow{\mathbf{p}}_{m,n}\cdot\overrightarrow{\bm{\omega}}}{||\overrightarrow{\bm{\omega}}||},
\end{IEEEeqnarray}
where $\overrightarrow{\mathbf{p}}_{m,n}$ represents the vector from the array center to the $(m,n)$-th antenna element.
The complete calibration factor $\Delta\mathbf{D}$ is constructed by extending the individual calibration terms into a matrix format:
\begin{IEEEeqnarray}{CCC}
	\Delta \mathbf{D}  =  [\Delta \mathbf{d}, \Delta \mathbf{d}, ..., \Delta \mathbf{d}]_{N_u\times N_b},\IEEEyesnumber\IEEEyessubnumber*\\
	\Delta \mathbf{d}  =  [\Delta d_{1,1},...,\Delta d_{1,N},...,\Delta d_{m,n},...,\Delta d_{M,N}]^T.
\end{IEEEeqnarray}
{With this spatial calibration, the free-space attenuation and phase terms are extended from scalar to matrix form via element-wise operations, given by
\begin{IEEEeqnarray}{rCl}
	\Delta \mathbf{A}_{ij} & = & \frac{c}{4\pi f}\Bigl(\mathbf{J}\oslash(\Delta\mathbf{D}_{ij} + \mathbf{J}\sum_{k=1}^{j-1}\delta_{ik})\Bigr),\IEEEyesnumber\IEEEyessubnumber*\\
	\Delta \bm{\phi}_{ij} & = & -\frac{2\pi f}{c}(\Delta\mathbf{D}_{ij} + \mathbf{J}\sum_{k=1}^{j-1}\delta_{ik}),
\end{IEEEeqnarray}
where $\oslash$ denotes element-wise division operation.
Consequently, the accumulated attenuation factor in Eq. (\ref{T_f}) is updated as
\begin{IEEEeqnarray}{rCl}\label{T_final}
	\mathbf{T}_{ij}(f) & = & \Delta \mathbf{A}_{ij}\odot\exp(j\Delta \bm{\phi}_{ij}) \odot\prod_{k=1}^{j-1}\bm{\beta}_{ik}(f) \IEEEyesnumber\IEEEyessubnumber* \\
	& = & \Delta \mathbf{A}_{ij}\odot\exp(j\frac{-2\pi f}{c}\Delta\mathbf{D}_{ij})\IEEEnonumber*\\
	& & \odot\exp(j\frac{-2\pi f}{c}\mathbf{J}\sum_{k=1}^{j-1}\delta_{ik})\odot\prod_{k=1}^{j-1}\bm{\beta}_{ik}(f)\IEEEyessubnumber \\
	& = & \frac{c}{4\pi f}\cdot\exp(j\frac{-2\pi f}{c}\Delta\mathbf{D}_{ij})\oslash(\Delta\mathbf{D}_{ij}+\mathbf{J}\sum_{k=1}^{j-1}\delta_{ik})\IEEEnonumber\\
	& & \odot\prod_{k=1}^{j-1}\bm{\beta}_{ik}(f)e^{j\frac{-2\pi f\delta_{ik}}{c}},\IEEEyessubnumber 
\end{IEEEeqnarray}
where $\odot$ denotes element-wise multiplication operation, and the product symbol $\prod$ is applied element-wise across matrices.}
Finally, Eq. (\ref{basicFieldChannel}) can be rewritten as
\begin{IEEEeqnarray}{rCl}
	\hat{\mathbf{H}}^\mathrm{D}(f) = \sum_{i=1}^{N_a}\sum_{j=1}^{N_r}\bm{\alpha}_{ij}(f)\odot\mathbf{T}_{ij}(f)\odot\mathbf{C}_{ij}(f).\label{finalFormulation}
\end{IEEEeqnarray}
Additionally, the mapping function Eq. (\ref{basicFieldMapping}) is updated as:
{
\begin{IEEEeqnarray}{rCl}
	\mathcal{F}_{\bm{\Theta}}(\cdot): \Bigl(\mathbf{q}(f),\bm{\omega}\Bigr)\longrightarrow\Bigl(\mathbf{C}(f),\bm{\sigma}(f)\Bigr),\label{finalMapping}
\end{IEEEeqnarray}
where $\mathbf{q}(f)$ denotes the frequency-dependent radiator-specific query.}

\section{Construction of F$^4$-CKM} \label{methods}
In this section, we provide an in-depth elaboration on the proposed F$^4$-CKM and its implementation specifics.
\subsection{Construction Pipeline}

Our F$^4$-CKM is constructed by rendering the aforementioned MIMO-OFDM RF radiance field.
The overall construction pipeline, as illustrated in Fig. \ref{pipeline}, takes the uplink CSI as input and predicts the corresponding downlink CSI.
The normalized mean squared error (NMSE) loss $\mathcal{L}_{\mathrm{NMSE}}$ is adopted for NN optimization, given as
\begin{IEEEeqnarray}{rCl}
	\mathcal{L}_{\mathrm{NMSE}} = \frac{1}{N_s}\sum_{i=1}^{N_s}\frac{1}{N_c}\sum_{j=1}^{N_c}{\frac{||\mathbf{\hat{H}}^\mathrm{D}_{ij}-\mathbf{H}^\mathrm{D}_{ij}||^2}{||\mathbf{H}^\mathrm{D}_{ij}||^2}},
\end{IEEEeqnarray}
where $\mathbf{\hat{H}}^\mathrm{D}_{ij}$ and $\mathbf{H}^\mathrm{D}_{ij}$ represent the predicted and ground truth CSI of the $j$-th subcarrier in the $i$-th sample, respectively, and $N_s$ denotes the total number of data samples.
In particular, this pipeline comprises three major components:
\begin{itemize}
	\item \textbf{\textit{Radiator Sampling:}}
	This module aims to effectively capture the EM distribution within the scene, which involves angular and radial sampling.
	Specifically, we utilize the SFG algorithm to ensure a highly uniform spherical distribution of the angular sampling.
	\item \textbf{\textit{Guided Query Shaping:}}
	Serving as a data preprocessing module, this module augments the uplink CSI input for the WiRARE network.
	It leverages two auxiliary indicators to provide side information and guides the shaping of {queries} at each sampled wireless radiator.
	\item \textbf{\textit{WiRARE Network:}}
	This module learns the mapping $\mathcal{F}_{\bm{\Theta}}(\cdot)$ to represent the sampled wireless radiators.
	Its architecture is tailored to effectively capture the spatial-frequency correlations within the input.
\end{itemize}
In the following subsections, we elaborate on the design and implementation of each module.
\subsection{Radiator Sampling}

\begin{figure}[t]
	\centering
	\subfloat[]{\includegraphics[width=0.25\textwidth]{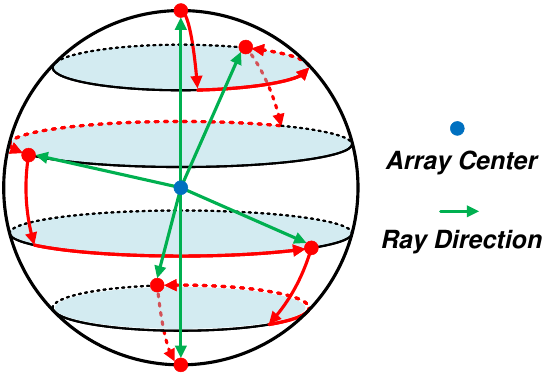}\label{SFGalgorithm}}
	$\,$
	\subfloat[]{\includegraphics[width=0.15\textwidth]{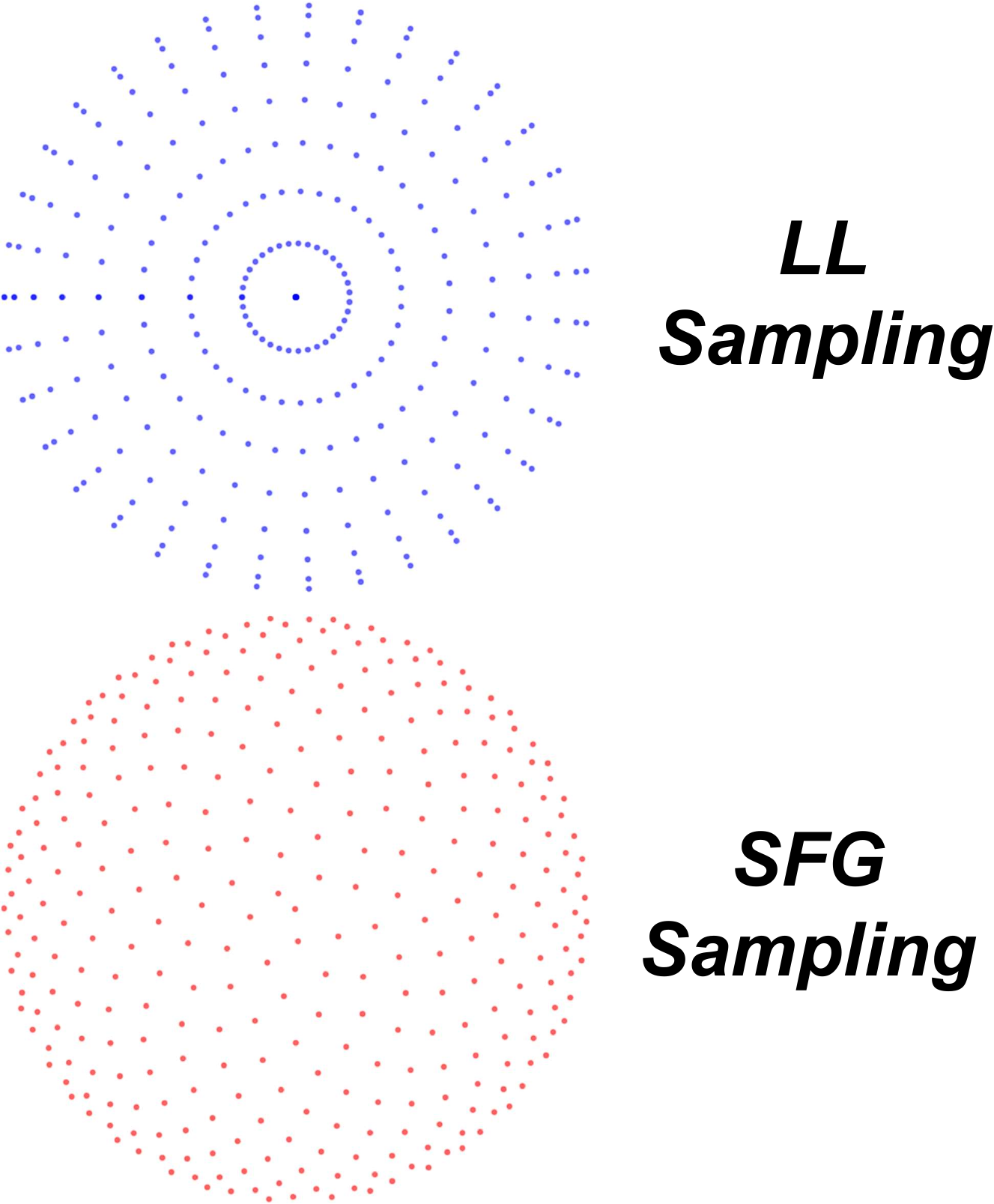}\label{SFGcompare}}
	\captionsetup{font=footnotesize}
	\caption{The SFG sampling algorithm. (a) illustrates the mechanisms of SFG sampling. (b) provides top views of $648$ sampled rays using LL sampling and SFG sampling, respectively. The angular resolution for LL sampling is set to $10^{\circ}$.}
	\label{SFG}
\end{figure}

Generally, there exists an inherent trade-off between sampling resolution and computational complexity.
While fine-grained sampling can naturally improve accuracy, it also increases computational overhead.
To achieve accurate yet low-complexity CKM construction, it is crucial to design a sampling strategy that enhances the sampling effectiveness with a given sampling resolution.
Furthermore, since the angular sampling process is responsible for capturing the AoAs of the received multipath RF signals, it has a more significant impact compared to the radial sampling process.
Therefore, we concentrate on the design of the angular sampling strategy while employing uniform sampling along the radial dimension.

Previous NeRF-based CKM frameworks have primarily employed two angular sampling strategies: latitude-longitude (LL) sampling \cite{Zhao2023nerf2} and AoA-assisted sampling \cite{Lu2024newrf}.
However, both methods have their limitations.
The LL sampling strategy, which divides the sphere into equally spaced grid points in latitude and longitude, is straightforward yet inherently flawed.
It leads to sparse sampling near the equator and dense sampling near the poles, thereby causing under-sampling and over-sampling issues, respectively.
In contrast, the AoA-assisted strategy attempts to enhance sampling efficiency by predicting AoAs at the UE's antenna array using NNs based on UE locations, and then performing radial sampling along these predicted directions.
However, this AoA prediction approach is highly dependent on the UE location inputs, which are subject to positioning errors and may raise privacy concerns, as discussed in Section \ref{Motivation}.
To overcome these limitations, we aim to sample rays that are uniformly distributed across a sphere, without relying on AoA information.

To achieve this target, the SFG algorithm \cite{Swinbank2006fibonacci} is employed for angular sampling, which enables evenly distributed sampling over the sphere and offers high configurability with respect to the number of rays.
Specifically, we consider tracing $K$ rays from the center of the UE's antenna array.
Our goal is to determine the directions of these $K$ rays in Cartesian coordinates.
By leveraging SFG, the direction of the $k$-th ray, $(x_k,y_k,z_k)$, can be expressed as
\begin{IEEEeqnarray}{rCl}
	z_k & = & 1 - \frac{2(k-1)}{K-1},\IEEEyesnumber\IEEEyessubnumber*\label{SFGz}\\
	x_k & = & \sqrt{1-z_n^2}\cdot\cos(2\pi k\Phi),\label{SFGx}\\
	y_k & = & \sqrt{1-z_n^2}\cdot\sin(2\pi k\Phi),\label{SFGy}
\end{IEEEeqnarray}
where $\Phi$ represents the golden ratio $(1+\sqrt{5})/2$.
Here, Eq. (\ref{SFGz}) divides the unit sphere into $K-1$ layers of equal thickness, with one sampling point (corresponding to one ray) per layer starting from the pole.
Thus, each layer has an equal surface area of $4\pi/N$, ensuring uniform sampling along the vertical axis.
Meanwhile, Eqs. (\ref{SFGx}) and (\ref{SFGy}) determine the longitude of each point using the golden angle $2\pi\Phi$, which effectively distributes new points into the largest angular gaps created by previous samples.
Fig. \ref{SFG}(a) illustrates the SFG-based sampling distribution, while Fig. \ref{SFG}(b) compares top-down views of $648$ rays sampled using the LL method and the SFG method.

\subsection{Guided Query Shaping}

\begin{figure}[t]
	\centering
	\includegraphics[width=0.4\textwidth]{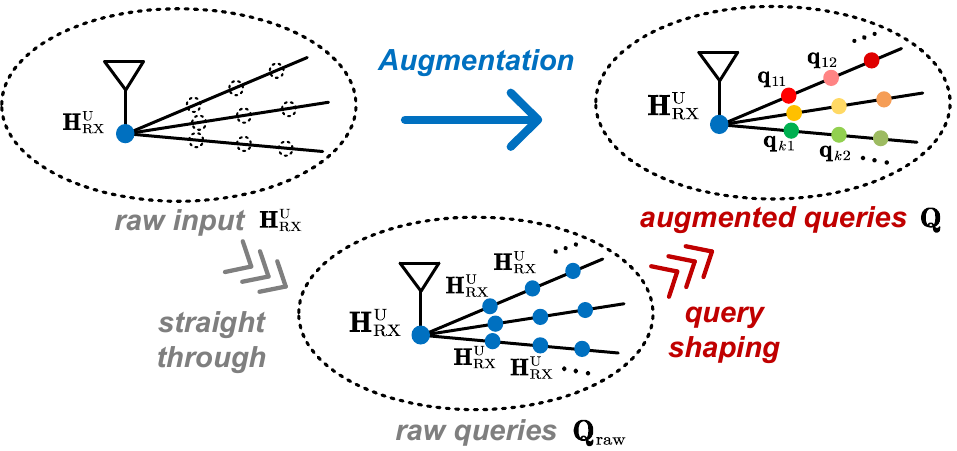}
	\captionsetup{font=footnotesize}
	\caption{The uplink CSI input augmentation process.}
	\label{augmentation}
\end{figure}

\begin{figure}[t]
	\centering
	\includegraphics[width=0.4\textwidth]{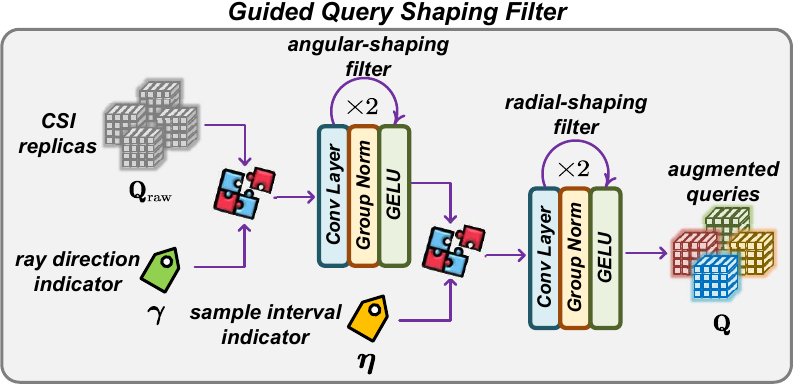}
	\captionsetup{font=footnotesize}
	\caption{The architecture of the guided shaping filter module.}
	\label{shapingFilter}
\end{figure}

After the radiator sampling stage, we obtain a set of rays $\{\bm{\omega}_i\}_{i=1}^{N_a}$ originating from the UE's antenna array and a series of wireless radiators sampled along each ray based on the sampling intervals $\{\delta_{ij}\}_{j=1}^{N_r}$.
{To learn the mapping in Eq. (\ref{finalMapping}), the radiator-specific queries $\{\mathbf{q}_{ij}(f)\}$ are required.
To bridge this gap, we construct the queries by jointly leveraging available information from two sources.
The first is the uplink CSI observed at the UE’s antenna array\footnote{In implementation, all complex-valued CSI data are converted into a real-valued format to facilitate NN training. Specifically, the real and imaginary parts of the CSI are concatenated along the frequency dimension.}, denoted as $\mathbf{H}^\mathrm{U}_{\mathrm{RX}}\in\mathbb{R}^{2N_c\times N_u \times N_b}$, which provides a global informative reference shared across all radiators.
The second is radiator-specific geometric information, namely the ray directions $\{\bm{\omega}_i\}_{i=1}^{N_a}$ and sampling intervals $\{\delta_{ij}\}_{j=1}^{N_r}$, which uniquely identify each radiator’s spatial distribution relative to the UE.
Their combination enables the generation of informative radiator-specific queries for the WiRARE network.

As illustrated in Fig. \ref{augmentation}, we begin with a straight-through initialization by replicating $\mathbf{H}^\mathrm{U}_{\mathrm{RX}}$ to each sampled radiator, yielding an initial query tensor $\mathbf{Q}_{\mathrm{raw}}\in\mathbb{R}^{N_a\times N_r\times 2N_c\times N_u \times N_b}$.
These CSI replicas serve as the initial input queries to the WiRARE network, enabling basic learning capability.
Subsequently, we introduce a novel query shaping process, implemented via a dedicated guided shaping filter module.
This module learns to fine-tune $\mathbf{Q}_{\mathrm{raw}}$ on its angular and radial dimensions by considering the spatial distribution of the radiators, thereby achieving a radiator-specific discrimination.
Specifically, we design the shaping filter as a hierarchical module, which operates in two stages with the guidance of two indicators, as illustrated in Fig. \ref{shapingFilter}.}

\begin{figure*}[t]
	\centering
	\includegraphics[width=0.8\textwidth]{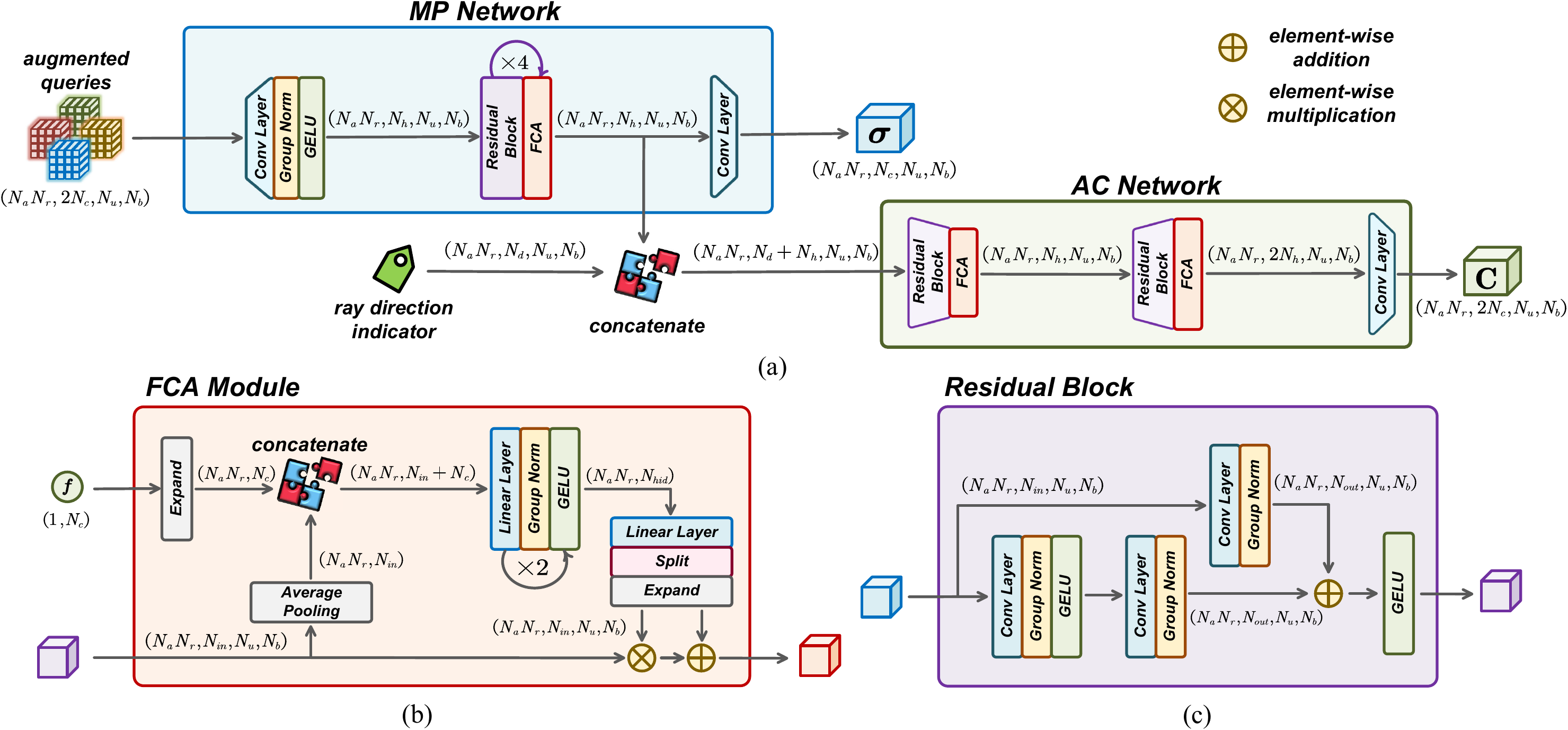}
	\captionsetup{font=footnotesize}
	\caption{The architecture of the WiRARE network. (a) provides an overview of WiRARE's entire architecture, which is built upon 2D convolutional layers. (b) and (c) depict the inner structures of the FCA modules and residual blocks within the WiRARE network, respectively.}
	\label{WiRARE}
\end{figure*}

In the first stage, the shaping filter fine-tunes {$\mathbf{Q}_{\mathrm{raw}}$} along its angular dimension.
Specifically, the ray directions $\{\bm{\omega}_i\}_{i=1}^{N_a}$ are first converted into an indicator $\bm{\gamma}$, which is processed through dimension expansion and tensor reshaping, given as
\begin{IEEEeqnarray}{rCl}
	\bm{\gamma}\in\mathbb{R}^{N_a\times 3} & \xrightarrow{\text{expand}} & \bm{\gamma}\in\mathbb{R}^{N_a\times N_r\times 3\times N_u \times N_b}\IEEEyesnumber\IEEEyessubnumber*\\
	& \xrightarrow{\text{reshape}} & \bm{\gamma}\in\mathbb{R}^{N_a N_r\times 3\times N_u \times N_b}.
\end{IEEEeqnarray}
Subsequently, the {query tensor $\mathbf{Q}_{\mathrm{raw}}$} is reshaped and concatenated with the indicator along the second dimension:
{\begin{IEEEeqnarray}{rLl}
	\mathbf{Q}_{\mathrm{raw}} & \xrightarrow{\text{reshape}} \mathbf{Q}_{\mathrm{raw}}\in\mathbb{R}^{N_a N_r\times 2N_c\times N_u \times N_b} & \IEEEyesnumber\IEEEyessubnumber*\\
	& \xrightarrow{\text{concatenate}} \mathbf{A}\in\mathbb{R}^{N_a N_r\times (2N_c+3)\times N_u \times N_b}, & \\
	& \mathbf{A}=\mathbf{Q}_{\mathrm{raw}}\oplus_2\bm{\gamma}, &
\end{IEEEeqnarray}}\noindent
where $\mathbf{A}$ denotes the concatenated tensor, and $\oplus_2$ represents the tensor concatenation operation along the second dimension.
Next, $\mathbf{A}$ is passed through the angular-shaping layer $\mathcal{G}_{\bm{\theta}}(\cdot)$, which is designed based on 2D convolutional layers:
{\begin{IEEEeqnarray}{rCl}
	\mathbf{Q}_{\mathrm{ang}}=\mathcal{G}_{\bm{\theta}}(\mathbf{A})\in\mathbb{R}^{N_a N_r\times 2N_c\times N_u \times N_b},
\end{IEEEeqnarray}
where $\mathbf{Q}_{\mathrm{ang}}$ represents the query tensor fine-tuned on the angular dimension, and $\bm{\theta}$ denotes the trainable parameters of the angular-shaping layer.}

In the second stage, the shaping filter refines {$\mathbf{Q}_{\mathrm{ang}}$} along its radial dimension.
Similar to the first stage, the sampling intervals $\{\delta_{ij}\}_{j=1}^{N_r}$ are transformed into an indicator $\bm{\eta}$, given as
\begin{IEEEeqnarray}{rCl}
	\bm{\eta}\in\mathbb{R}^{N_r} & \xrightarrow{\text{expand}} & \bm{\eta}\in\mathbb{R}^{N_a\times N_r\times 2N_c\times N_u \times N_b}\IEEEyesnumber\IEEEyessubnumber*\\
	& \xrightarrow{\text{reshape}} & \bm{\eta}\in\mathbb{R}^{N_a\times N_r\times 2N_c\times N_u N_b}.
\end{IEEEeqnarray}
Following this, {$\mathbf{Q}_{\mathrm{ang}}$} is reshaped and concatenated with $\bm{\eta}$ along the second dimension, given as
{\begin{IEEEeqnarray}{rLl}
	\mathbf{Q}_{\mathrm{ang}} & \xrightarrow{\text{reshape}} \mathbf{Q}_{\mathrm{ang}}\in\mathbb{R}^{N_a\times N_r\times 2N_c\times N_u N_b} & \IEEEyesnumber\IEEEyessubnumber*\\
	& \xrightarrow{\text{concatenate}} \mathbf{B}\in\mathbb{R}^{N_a\times 2N_r\times 2N_c\times N_u N_b}, & \\
	& \mathbf{B}=\mathbf{Q}_{\mathrm{ang}}\oplus_2\bm{\eta}, &
\end{IEEEeqnarray}}\noindent
where $\mathbf{B}$ represents the concatenated tensor.
Subsequently, $\mathbf{B}$ is processed through the radial-shaping layer $\mathcal{G}_{\bm{\phi}}(\cdot)$, given as
{\begin{IEEEeqnarray}{rCl}
	\mathbf{Q}_{\mathrm{rad}} & =\mathcal{G}_{\bm{\phi}}(\mathbf{B})\in\mathbb{R}^{N_a\times N_r\times 2N_c\times N_u N_b}, & \IEEEyesnumber\IEEEyessubnumber*\\
	\mathbf{Q}_{\mathrm{rad}} & \xrightarrow{\text{reshape}} \mathbf{Q}\in\mathbb{R}^{N_a N_r\times 2N_c\times N_u \times N_b}, &
\end{IEEEeqnarray}
where $\mathbf{Q}_{\mathrm{rad}}$ denotes the queries fine-tuned along the radial dimension, $\bm{\phi}$ are the trainable parameters of the radial-shaping filter, and $\mathbf{Q}$ represents the final output of the guided shaping filter module, which is then fed into the WiRARE network.}

In summary, the entire guided {query shaping} process can be concisely represented by the following expression:
{\begin{IEEEeqnarray}{rCl}
	\mathbf{Q} = \mathcal{G}_{\bm{\phi}}(\mathcal{G}_{\bm{\theta}}(\mathbf{Q}_{\mathrm{raw}}\oplus_2\bm{\gamma})\oplus_2\bm{\eta}),
\end{IEEEeqnarray}}\noindent
where the operations of dimension expansion and tensor reshaping are implicitly incorporated.
The trainable parameters $\{\bm{\theta},\bm{\phi}\}$ of the guided {shaping filter} are jointly optimized with the WiRARE network in an end-to-end manner, as described in the following subsection.

\subsection{WiRARE Network}


With the {augmented queries $\mathbf{Q}$} prepared, the WiRARE network is designed to predict the material properties and aggregating coefficients of all the sampled wireless radiators.
The network architecture is presented in Fig. \ref{WiRARE}, which features a spatial-aware backbone and frequency-aware fine-tuning through the FCA modules.
\subsubsection{Spatial-Aware Backbone}
Existing methods \cite{Zhao2023nerf2,Lu2024newrf,Wen2024wrf} typically implement their scene representation NNs based on fully-connected layers.
While these structures are sufficient for learning basic RF radiance fields, they fail to effectively leverage the spatial correlations within the data, thereby limiting the networks' capacity to learn more complex RF radiance fields in multiple antenna systems.
In contrast, our WiRARE network is constructed using 2D convolutional layers, which are particularly effective at capturing spatial correlations within the data.
This spatial awareness significantly enhances WiRARE's capacity for MIMO RF radiance field rendering.

Specifically, WiRARE consists of two sub-networks: the material property (MP) network and the aggregating coefficient (AC) network, tasked with predicting the material properties $\bm{\sigma}$ and aggregating coefficients $\mathbf{C}$, respectively.
The input {$\mathbf{Q}$} is initially processed by the MP network, which contains an up-sampling convolutional layer that includes $N_h$ hidden channels, a group normalization (GN) layer \cite{Wu2018group}, and a GELU activation function \cite{Hendrycks2016gaussian}.
Subsequent to these initial layers, the output is passed through four residual blocks, the inner structure of which is depicted in Fig. \ref{WiRARE}(c).
Specifically, the main branch of each residual block consists of two convolutional layers, combined with GN layers and GELU activation.
In addition, a shortcut connection is added to the output of the second convolution, which is implemented either as (i) an identity mapping (when hidden dimensions are preserved) or as (ii) a $1 \times 1$ convolution (when down-sampling is applied).
The sum of the main and shortcut branches is then passed through a final GELU activation before being forwarded to subsequent layers.
The output features from these blocks then diverge into two distinct processing flows.
The first flow involves a down-sampling convolutional layer that produces the prediction of material properties $\bm{\sigma}$.
The second flow directs the features into the {AC network} for predicting the aggregating coefficients $\mathbf{C}$.
Importantly, we model the wireless radiators as anisotropic sources that radiate direction-dependent RF signals, which enhances the network’s ability to capture multipath propagation effects.
To facilitate the learning of these directional characteristics, the output features from the MP network are concatenated with the ray direction indicator $\bm{\gamma}$ before being fed into the {AC network}.
Regarding the {AC network}, the input features first pass through a down-sampling residual block, succeeded by an up-sampling residual block and a down-sampling convolutional layer.
Finally, the predicted aggregating coefficients $\mathbf{C}$ are generated in a real-valued format, with {their} real and imaginary parts concatenated along the frequency dimension.
The predicted $\bm{\sigma}$ and $\mathbf{C}$ are then leveraged to obtain the final prediction of the downlink CSI $\hat{\mathbf{H}}^\mathrm{D}$ based on Eq. (\ref{finalFormulation}).

\subsubsection{Frequency-Aware Fine-Tuning}

To further enhance WiRARE's prediction performance beyond its spatial-aware backbone design, we introduce frequency-aware fine-tuning by developing the FCA module.
This module serves as a feature modulator that refines intermediate features by incorporating explicit frequency values of each subcarrier $f$ (GHz) as side information.

Specifically, the FCA module fine-tunes the input features via affine transformations.
The weights and biases for these transformations are jointly determined by both the input features and the frequency values of each subcarrier.
As presented in Fig. \ref{WiRARE}(b), the input features and frequency values are first processed by average pooling and dimension expansion, respectively, to align their dimensions and enable the merging of these two streams of information.
Subsequently, three fully-connected layers, combined with GN layers and GELU activations, are employed to process the merged information stream.
Finally, the weights and biases for the final affine transformation on the input features are generated from the processed information.
It is important to note that both $\bm{\sigma}$ and $\mathbf{C}$ are frequency-dependent.
Therefore, the FCA modules are embedded into both the MP network and the {AC network}, attached to the tail of the residual blocks, as illustrated in Fig. \ref{WiRARE}(a).

\section{Simulation Results} \label{Simulation}
In this section, we evaluate the performance of our F$^4$-CKM through extensive experiments.
We incorporate a variety of metrics and advanced baselines, conducting performance comparisons across both simulated and real-world MIMO-OFDM channel datasets.
\subsection{Simulation Setup}
\subsubsection{Datasets}

\begin{figure}[t]
	\centering
	\includegraphics[width=0.45\textwidth]{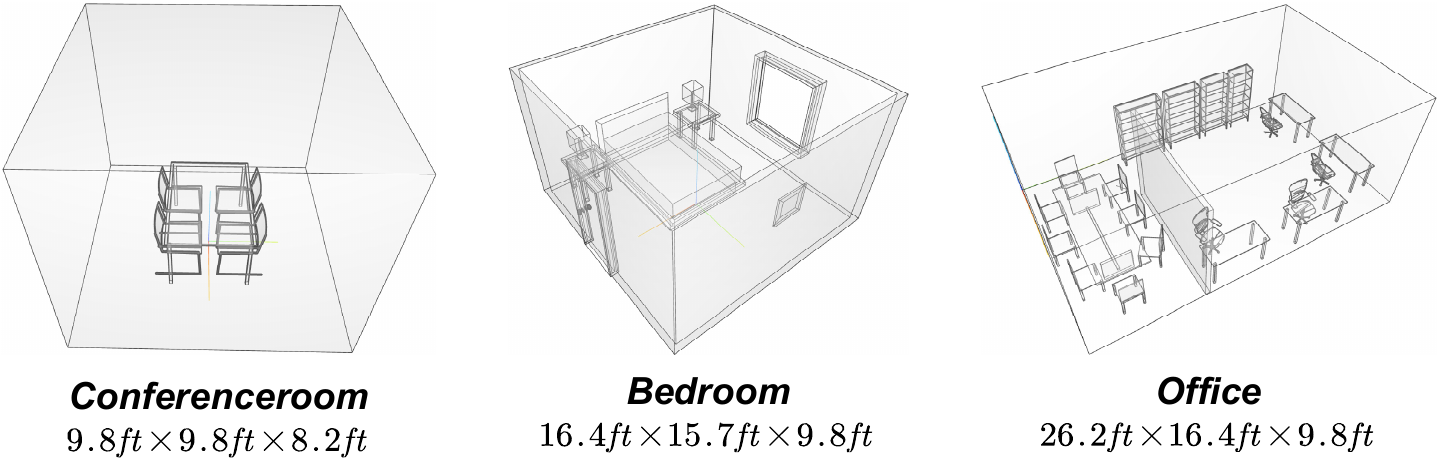}
	\captionsetup{font=footnotesize}
	\caption{The geometry of three indoor 3D environment models.}
	\label{3Dmodels}
\end{figure}

\renewcommand{\arraystretch}{1.5}
\begin{table}[t]\scriptsize
	\centering
	\caption{Parameter settings for simulated dataset generation.}
	\label{simDataset}
	\begin{tabular}{c|c|c|c}
		\toprule
		\multirow{2}{*}{\textbf{Environment}} & \multicolumn{2}{c|}{\textbf{Parameters}} & \multirow{2}{*}{\textbf{Size (train$|$test)}} \\
		\cline{2-3}
		& TX Coordinate (m) & Band (Up$|$Down) & \\
		\hline
		\multirow{3}{*}{Conferenceroom} & $[-1.2, 0.0, 1.7]$ & $2.415|2.465$ GHz & $9,863 | 1,976$ \\
		& $[-1.2, 0.0, 1.7]$ & $3.515|3.565$ GHz & $9,869 | 1,973$  \\
		& $[-1.2, 0.0, 1.7]$ & $6.715|6.765$ GHz & $9,889 | 1,976$ \\
		& {$[-1.2, 0.0, 1.7]$} & {$28.015|28.065$ GHz} & {$9,922 | 1,983$} \\
		\hline
		Bedroom & $[-1.2, 0.0, 2.7]$ & $2.415|2.465$ GHz & $17,824 | 3,589$ \\
		\hline
		Office & $[0.3, 3.7, 2.7]$ & $2.415|2.465$ GHz & $28,082 | 5,633$ \\
		\bottomrule
	\end{tabular}
\end{table}

\begin{figure*}[t]
	\centering
	\includegraphics[width=0.9\textwidth]{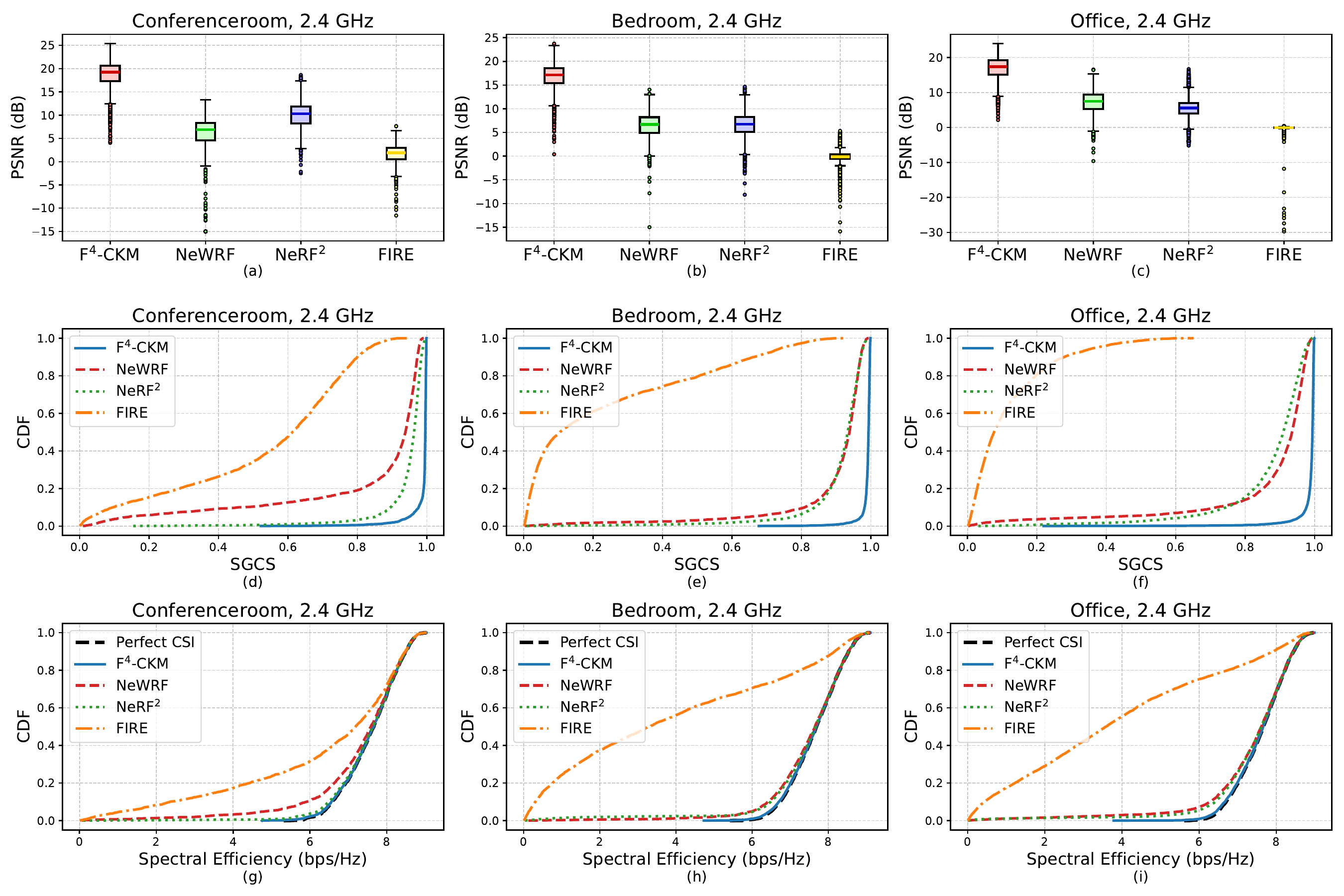}
	\captionsetup{font=footnotesize}
	\caption{The PSNR, SGCS, and spectral efficiency performance in different indoor environments.}
	\label{sim_envs}
\end{figure*}

\begin{itemize}
	\item \textbf{\textit{Simulated Dataset.}}
	The simulated dataset is generated using the MATLAB platform.
	Three indoor 3D environment models referenced in \cite{Lu2024newrf} are employed for dataset generation.
	The geometry and physical dimensions of these environments are shown in Fig. \ref{3Dmodels}.
	In each environment, a fixed transmitter equipped with a $4\times 4$ UPA and multiple receivers randomly distributed throughout the scene, each with a $2\times 2$ UPA, are set up.
	The OFDM bandwidth is configured at $20$ MHz for both uplink and downlink channels, divided into $64$ subcarriers with each spaced $312.5$ KHz from its adjacent subcarriers.
	In our experiments, $52$ out of these $64$ subcarriers are utilized for channel measurements.
	The frequency band gap between uplink and downlink channels is set to $50$ MHz.
	We adopt the shooting-and-bouncing ray tracing (SBR) method to obtain simulated uplink and downlink channel measurements for each receiver.
	{Datasets are constructed for each environment at the $2.4$ GHz band, with three additional datasets produced for the conference room scenario at $3.5$ GHz, $6.7$ GHz, and $28.0$ GHz bands, respectively.}
	Comprehensive parameter settings for the simulated dataset generation are detailed in Table \ref{simDataset}.
	\item \textbf{\textit{Real-World Dataset.}}
	In addition, the public Argos channel dataset \cite{Shepard2016understanding} is incorporated for performance evaluation.
	This dataset comprises real-world MIMO-OFDM channel measurements collected across various environments, where the BS is equipped with $104$ antennas and serves up to $8$ UEs.
	The OFDM bandwidth is set to $20$ MHz, divided into $64$ subcarriers, with 52 utilized for channel measurement purposes.
	Consistent with previous works \cite{Zhao2023nerf2,Wen2024wrf}, we designate the first $26$ subcarriers to the uplink and the remaining $26$ subcarriers to the downlink.
	
\end{itemize}

\begin{figure*}[t]
	\centering
	\includegraphics[width=0.9\textwidth]{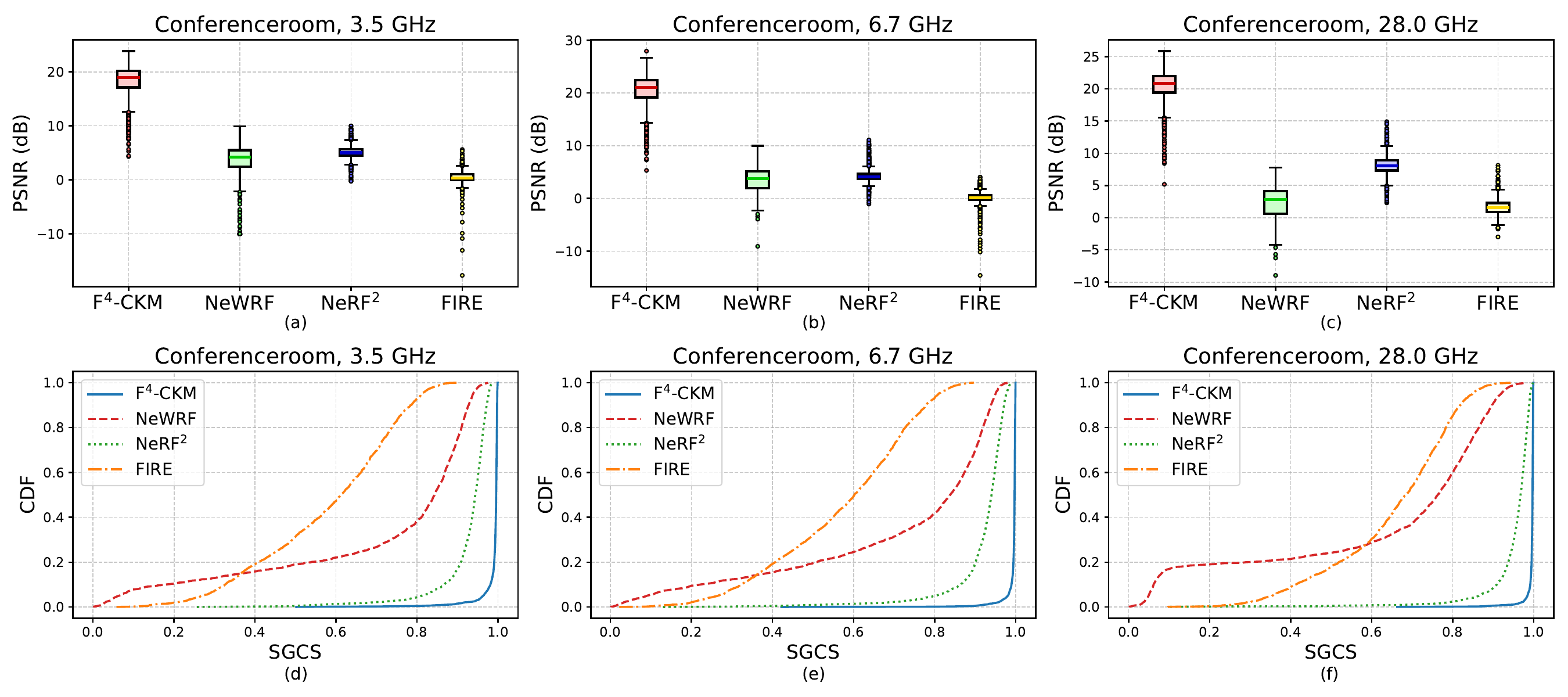}
	\captionsetup{font=footnotesize}
	\caption{The PSNR and SGCS performances at different frequency bands in the conference room environment. For each frequency band, all algorithms are trained and evaluated at that same band.}
	\label{sim_frequencies}
\end{figure*}

\subsubsection{Metrics}
{We employ three metrics to evaluate the accuracy of channel prediction, referred to as prediction signal-to-noise ratio (PSNR), square generalized cosine similarity (SGCS), and spectral efficiency, respectively.}
The PSNR is defined as
\begin{IEEEeqnarray}{rCl}
	\mathrm{PSNR}\,\,(\mathrm{dB}) = -10\log_{10}(\frac{||\mathbf{\hat{H}}^\mathrm{D}-\mathbf{H}^\mathrm{D}||^2}{||\mathbf{H}^\mathrm{D}||^2}),
\end{IEEEeqnarray}
where $\hat{\mathbf{H}}^\mathrm{D}$ and $\mathbf{H}^\mathrm{D}$ represent the predicted and actual downlink CSI, respectively.
A higher PSNR indicates a more accurate prediction.
Additionally, the SGCS metric, aimed at evaluating the angular domain accuracy of CSI predictions, is given by
\begin{IEEEeqnarray}{rCl}
	\mathrm{SGCS} = \frac{1}{N_c}\sum_{i=1}^{N_c}{\frac{||\mathbf{w}_{i}^{H}\hat{\mathbf{w}}_i||^2}{||\mathbf{w}_i||^2||\hat{\mathbf{w}}_i||^2}},
\end{IEEEeqnarray}
where $\hat{\mathbf{w}}$ and $\mathbf{w}$ denote the ideal eigenvectors derived from the predicted and actual CSI matrices, respectively.
A higher SGCS value signifies a more accurate prediction of CSI in the angular domain, which is beneficial for beamforming in MIMO systems.
{Moreover, the spectral efficiency is used to evaluate the end-to-end system performance.
We assume that both BS and UE employ discrete Fourier transform codebooks and the signal-to-noise ratio is $\gamma = 10$ dB, then the spectral efficiency is computed as
\begin{IEEEeqnarray}{rCl}
	R\,\,(\mathrm{bps/Hz}) = \frac{1}{N_c}\sum_f\log_2\Bigl(1+\gamma\cdot||\mathbf{w}^H_*\mathbf{H}^{\mathrm{D}}(f)\mathbf{f}_*||^2\Bigr),\IEEEeqnarraynumspace
\end{IEEEeqnarray}
where $\mathbf{f}_*\in\mathbb{C}^{N_b}$ and $\mathbf{w}_*\in\mathbb{C}^{N_u}$ denote the transmit and receive beamforming vectors, respectively, which are chosen to maximize the channel gain based on the predicted downlink CSI $\hat{\mathbf{H}}^\mathrm{D}(f)$, given by
\begin{IEEEeqnarray}{rCl}
	(\mathbf{w}_*,\mathbf{f}_*) = \arg \max_{\mathbf{f}\in\mathcal{F},\mathbf{w}\in\mathcal{W}} \frac{1}{N_c}\sum_{f}||\mathbf{w}^H\hat{\mathbf{H}}^{\mathrm{D}}(f)\mathbf{f}||^2,
\end{IEEEeqnarray}
where $\mathcal{F}$ and $\mathcal{W}$ are the codebooks at the BS and UE, respectively.}

\subsubsection{Baselines}
We compare F$^4$-CKM with four baselines.
\begin{itemize}
	\item \textbf{NeWRF} \cite{Lu2024newrf}: 
	NeWRF is a NeRF-based approach that predicts CSI using user locations.
	We use the open-source code of this method and train it on our datasets.
	Since their code is designed for single-input single-output (SISO) systems, we adapt the NeWRF approach to our MIMO datasets by collapsing the antenna dimension into the data sample dimension and performing element-wise CSI predictions.
	\item \textbf{NeRF$^2$} \cite{Zhao2023nerf2}: 
	NeRF$^2$ is a NeRF-based approach that predicts downlink CSI based on uplink CSI.
	We utilize the open-source code provided by the authors.
	\item \textbf{FIRE} \cite{Liu2021fire}: 
	FIRE is a model-free method based on the variational autoencoder (VAE) architecture and predicts downlink CSI from uplink CSI.
	We utilize the open-source code available for this method.
\end{itemize}
\subsubsection{Implementation Details}


We implement our F$^4$-CKM and all baselines using the PyTorch platform, a $2.60$ GHz Intel(R) Xeon(R) Platinum 8350C CPU, and an NVIDIA GeForce RTX 4090 GPU.
To train F$^4$-CKM, the Adam optimizer \cite{Kingma2014adam} with an initial learning rate of $5\times 10^{-5}$ and a batch size of $8$ is employed.
The learning rate is adjusted over time using the ReduceLROnPlateau scheduler with a patience of $10$ and a reduction factor of $0.9$.
The number of sampled rays and radiators sampled along each ray is set to $24$ and $32$, respectively, with the radial sampling range set to $9$ m.
{The code is available at: \url{https://github.com/kqzzzz/F4CKM}.}

\subsection{Evaluation over Simulated Datasets}

{In Fig. \ref{sim_envs}, we compare F$^4$-CKM with the baselines across three distinct indoor environments.}
Specifically, Figs. \ref{sim_envs}(a), \ref{sim_envs}(b), and \ref{sim_envs}(c) depict the PSNR performance in the conference room, the bedroom, and the office, respectively.
The median PSNR values achieved by F$^4$-CKM in these environments are $19.23$ dB, $17.13$ dB, and $17.39$ dB, respectively.
In contrast, the highest PSNR values among the baselines are $9.71$ dB, $6.67$ dB, and $7.48$ dB, respectively.
Furthermore, Figs. \ref{sim_envs}(d), \ref{sim_envs}(e), and \ref{sim_envs}(f) present the SGCS performance through cumulative distribution function (CDF) curves.
Notably, the CDF curve of F$^4$-CKM exhibits an almost vertical increase near the point where SGCS approaches $1.0$.
In particular, over $90\%$ of the results from F$^4$-CKM achieve an SGCS value higher than $0.976$ across all three environments.
This remarkable consistency demonstrates the reliability of F$^4$-CKM in applications related to beamforming.

{Additionally, results of system-level performance evaluation are presented in Figs. \ref{sim_envs}(g), \ref{sim_envs}(h), and \ref{sim_envs}(i), where the “Perfect CSI” curves correspond to the spectral efficiency upper bound assuming error-free downlink CSI acquisition.
Notably, the CDF curve of F$^4$-CKM closely approaches that of the “Perfect CSI”, indicating highly reliable channel prediction.
Specifically, F$^4$-CKM achieves spectral efficiencies exceeding $4.7$ bps/Hz (conference room and bedroom) and $3.7$ bps/Hz (office) for all samples in the test set, whereas the baselines exhibit significant performance variability, suggesting occasional prediction failure.}
It is readily seen that F$^4$-CKM significantly outperforms all baselines in all three environments.
This superiority is attributed to the spatial-frequency-aware system formulation and network design of F$^4$-CKM, which effectively captures the characteristics of wireless channels.

Fig. \ref{sim_frequencies} shows the performance of F$^4$-CKM across different frequency bands in the conference room environment.
It is evident that F$^4$-CKM maintains a robust PSNR performance of approximately $20$ dB and consistently achieves SGCS values that are closely concentrated around $1.0$, outperforming all baselines.
This robustness stems from the frequency-aware system formulation of F$^4$-CKM, as well as the design of the FCA module.
By incorporating frequency values as side information, the FCA module enables F$^4$-CKM to achieve consistent performance across a broad spectrum.

\begin{figure}[t]
	\centering
	\includegraphics[width=0.49\textwidth]{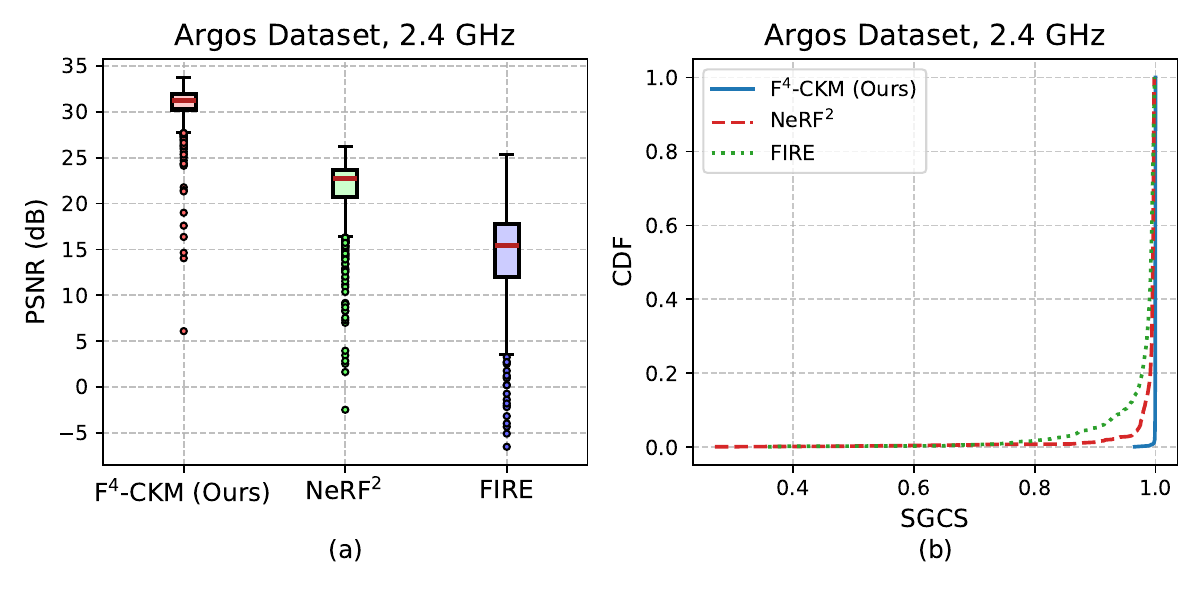}
	\captionsetup{font=footnotesize}
	\caption{The PSNR and SGCS performance over the Argos dataset.}
	\label{sim_argos}
\end{figure}

\renewcommand{\arraystretch}{1.5}
\begin{table}[t]\scriptsize
	\centering
	\caption{PSNR Performance comparison over the Argos dataset.}
	\label{argosCompare}
	\begin{tabular}{c|c|c|c|c|c}
		\toprule
		& \makecell{upper\\whisker} & \makecell{upper\\quartile} & median & \makecell{lower\\quartile} & \makecell{lower\\whisker} \\
		\midrule
		\midrule
		F$^4$-CKM & $34.30$ dB & $31.85$ dB & $31.19$ dB & $30.22$ dB & $27.77$ dB \\
		WRF-GS+ & $29.00$ dB & $25.51$ dB & $23.91$ dB & $20.88$ dB & $14.40$ dB \\
		WRF-GS & $26.73$ dB & $24.35$ dB & $22.98$ dB & $20.83$ dB & $15.65$ dB \\
		NeRF$^2$ & $26.80$ dB & $22.26$ dB & $21.17$ dB & $19.24$ dB & $14.70$ dB \\
		FIRE & $26.46$ dB & $17.79$ dB & $15.44$ dB & $12.01$ dB & $3.34$ dB \\
		R2F2 & $11.54$ dB & $9.63$ dB & $8.57$ dB & $6.82$ dB & $2.65$ dB \\
		OptML & $11.07$ dB & $9.37$ dB & $8.47$ dB & $6.10$ dB & $1.25$ dB \\
		\bottomrule
	\end{tabular}
\end{table}

\subsection{Evaluation over Real-World Datasets} 

\begin{figure}[t]
	\centering
	\includegraphics[width=0.45\textwidth]{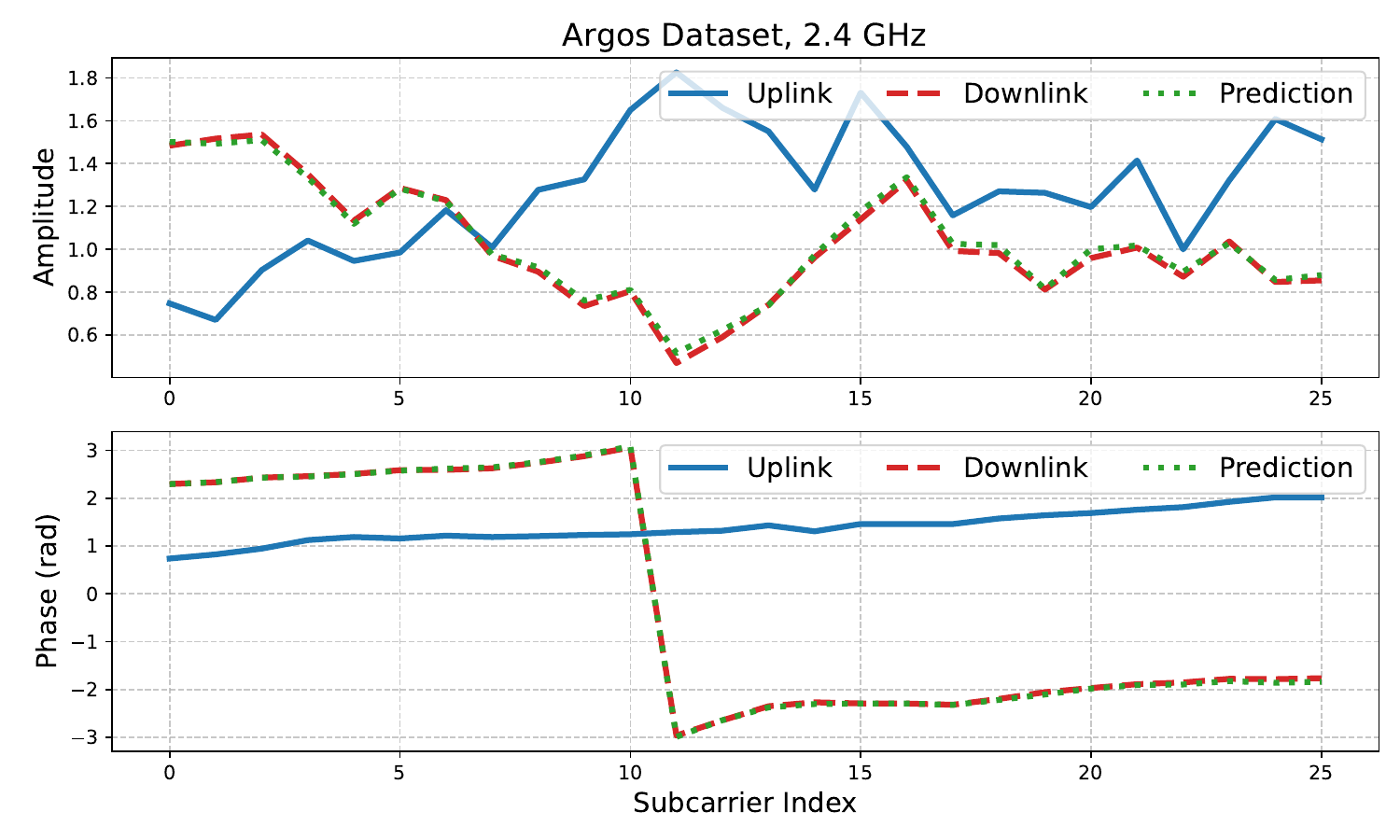}
	\captionsetup{font=footnotesize}
	\caption{Comparison of the ground truth and predicted downlink CSI using F$^4$-CKM in terms of amplitude and phase versus subcarrier indices.}
	\label{sim_example}
\end{figure}

To further demonstrate the practical effectiveness of our approach, we evaluate the performance of F$^4$-CKM using the Argos real-world dataset \cite{Shepard2016understanding}.
Specifically, Fig. \ref{sim_argos} presents a detailed comparison of the PSNR and SGCS performances between F$^4$-CKM, NeRF$^2$, and FIRE using the Argos dataset.
From Fig. \ref{sim_argos}(a), we can observe that F$^4$-CKM clearly outperforms all baselines, achieving a median PSNR value exceeding $30$ dB.
Moreover, the box plot of F$^4$-CKM is notably flatter than those of NeRF$^2$ and FIRE, which indicates that F$^4$-CKM delivers more robust and reliable performance.
Additionally, Fig. \ref{sim_argos}(b) shows that F$^4$-CKM achieves SGCS scores higher than $0.965$ for $100\%$ of the test data.
This result further underscores F$^4$-CKM's remarkable performance in channel prediction.
To provide a more intuitive illustration of the results, Fig. \ref{sim_example} selects the sample with the median PSNR value and presents a comparison between the ground truth and the predicted downlink CSI.
This is achieved by extracting a slice along the antenna dimension.
We decompose the complex-valued CSI into its amplitude and phase components and plot these values versus subcarrier indices.
It is observed that the predicted and ground truth curves almost overlap with each other, which highlights the exceptional accuracy of F$^4$-CKM in channel prediction tasks.

Table \ref{argosCompare} elaborates on the PSNR results depicted in Fig. \ref{sim_argos} and extends the comparison to four additional baselines, namely WRF-GS \cite{Wen2024wrf}, WRF-GS+ \cite{Wen2024wrf}, R2F2 \cite{Vasisht2016eliminating}, and OptML \cite{Bakshi2019fast}.
\begin{itemize}
	\item \textbf{R2F2}: This model-based approach uses uplink CSI to infer downlink CSI by solving an optimization problem to determine the multipath information for prediction.
	\item \textbf{OptML}: OptML utilizes deep learning techniques to extract multipath information from uplink CSI, which is subsequently exploited to infer downlink CSI.
	\item \textbf{WRF-GS}: WRF-GS combines the 3DGS approach with NNs to reconstruct RF radiance fields, thereby predicting downlink CSI from uplink CSI.
	\item \textbf{WRF-GS+}: An enhanced version of WRF-GS, which employs deformable 3D Gaussians to model both static and dynamic components of RF radiance fields.
\end{itemize}
To ensure fairness and consistency, the results for these four baseline methods are drawn from \cite{Wen2024wrf}, where they were evaluated using the same dataset.
From this table, our F$^4$-CKM approach achieves a remarkable performance leap over all baseline methods.
Notably, F$^4$-CKM attains the best performance with a median PSNR value of $31.19$ dB, which is $7.28$ dB higher than the SOTA method, WRF-GS+.

\renewcommand{\arraystretch}{1.2}
\begin{table}[t]\scriptsize
	\centering
	\caption{Latency and computational overhead comparison in Argos dataset}  
	\label{complexity_argos}
	\begin{tabular}{lcccc}
		\toprule
		\textbf{Model} & \makecell{\textbf{Latency} \\ \textbf{(ms)}} & \textbf{GFLOPs} & \makecell{\textbf{Params} \\ \textbf{(M)}} & \makecell{\textbf{Mean PSNR} \\ \textbf{(dB)}} \\
		\midrule
		F$^4$-CKM & $5.756$  & $37.27$ & $3.41$ & $30.74$ \\
		F$^4$-CKM-Lite & $4.326$  & $6.34$ & $0.51$ & $27.40$ \\
		NeRF$^2$ & $8.775$  & $117.03$ & $0.71$ & $21.51$ \\
		FIRE & $0.460$  & $9.45\times 10^{-5}$ & $0.05$ & $14.47$ \\
		\bottomrule
	\end{tabular}
\end{table}

{\subsection{Complexity Analysis}

As the proposed F$^4$-CKM incurs additional computation, it is necessary to analyze the computational complexity for efficiency considerations.
The main cost stems from the WiRARE network and the radiator aggregation operation in Eq. (\ref{finalFormulation}).
\begin{itemize}
	\item \textbf{WiRARE Network:}
	The computational complexity of WiRARE is dominated by its residual blocks and FCA modules.
	Each FCA module incurs $\mathcal{O}\big(N_{\mathrm{hid}}^2 + N_{\mathrm{in}} N_{\mathrm{hid}} + N_c N_{\mathrm{hid}} + N_{\mathrm{in}} N_u N_b\big)$ floating-point operations (FLOPs), primarily from the fully-connected layers and the channel-wise affine transformation.
	Here, $N_{\mathrm{in}}$ denotes the channel dimension of the FCA module's input features and $N_{\mathrm{hid}}$ represents the hidden dimension.
	Each residual block contributes $\mathcal{O}\big(N_{\mathrm{out}} (N_{\mathrm{in}} + N_{\mathrm{out}}) N_u N_b\big)$ FLOPs, where $N_{\mathrm{in}}$ and $N_{\mathrm{out}}$ are the input and output channel counts, respectively (the $K^2$ factor is omitted since the convolutional kernel size $K$ is fixed at $3$).
	For the full WiRARE network with $L$ interleaved residual–FCA stages and assuming $N_{\mathrm{in}}, N_{\mathrm{out}} = \mathcal{O}(N_{\mathrm{hid}})$, the overall computational complexity scales as $\mathcal{O}\big(N_{\mathrm{hid}} (L N_{\mathrm{hid}} + N_c) N_u N_b \big)$.
	This result confirms that the computational complexity scales linearly with the MIMO array size, while $N_{\mathrm{hid}}$ and $L$ offer practical complexity control.
	\item \textbf{Radiator Aggregation:}
	The complexity of the radiator aggregation operation is dominated by the element-wise cumulative product in Eq. (\ref{T_final}).
	For each ray, computing the product of up to $j-1$ matrices $\bm{\beta}_{ij}(f) \in \mathbb{C}^{N_u \times N_b}$ for $j=1,\dots,N_r$ requires $\mathcal{O}(N_r^2 N_u N_b)$ operations.
	In contrast, the subsequent weighted summation in Eq.~(\ref{finalFormulation}) requires only $\mathcal{O}(N_r N_u N_b)$ operations per ray, making the total complexity being $\mathcal{O}(N_a N_r^2 N_c N_u N_b)$.
\end{itemize}
In summary, the overall complexity of F$^4$-CKM is
\begin{IEEEeqnarray}{rCl}
	\mathcal{F}_{\mathrm{total}} & = & \mathcal{O}\big(N_a N_r \cdot \big( N_{\mathrm{hid}} (L N_{\mathrm{hid}} + N_c)N_u N_b\big)\IEEEnonumber*\\
	& & + N_a N_r^2 N_c N_u N_b \big)\\
	& = & \mathcal{O}\big(N_a N_r N_u N_b \cdot \big( N_{\mathrm{hid}} (L N_{\mathrm{hid}} + N_c) + N_r N_c \big) \big),\IEEEyesnumber \IEEEeqnarraynumspace
\end{IEEEeqnarray}
which scales linearly with array size ($N_u N_b$), and is controllable via the number of rays $N_a$, the per-ray radiator count $N_r$, the network depth $L$, and the hidden dimension $N_{\mathrm{hid}}$.

To provide a quantitative complexity analysis, we measured the average inference latency, FLOPS, and model size of F$^4$-CKM.
For comparison, we also report the latency and computational overhead of representative baseline models.
Evaluations are conducted on the Argos dataset and the results are summarized in Table \ref{complexity_argos}.
All tests were conducted on a single GPU, with each metric averaged over $10$ independent runs across the full test set to mitigate measurement variance.
We can observe that F$^4$-CKM achieves lower inference latency and computational overhead than mainstream radiance field models such as NeRF$^2$, while delivering significantly higher CSI prediction fidelity.
According to 3GPP technical specification (TS) 38.331 for Release 18 \cite{3gpp_ts38331}, the measured inference latencies of F$^4$-CKM are compatible with commonly deployed CSI reporting periodicity configurations such as $10$ ms and $20$ ms.

Furthermore, we developed a lightweight version of F$^4$-CKM, named F$^4$-CKM-Lite, by reducing network depth from $6$ to $4$ and shrinking the hidden dimension from $128$ to $32$.
This lightweight variant cuts FLOPs by over $80\%$ with only marginal degradation in PSNR, while still outperforming all baselines.
These results confirm that our architecture enables scalable complexity control with only minor fidelity degradation, yielding a favorable complexity–accuracy trade-off.
Additionally, it is worth noting that established model compression and acceleration techniques can further reduce complexity and latency, which are regarded as orthogonal enhancements.}

\begin{figure}[t]
	\centering
	\begin{subfigure}[b]{0.45\textwidth}
		\includegraphics[width=\textwidth]{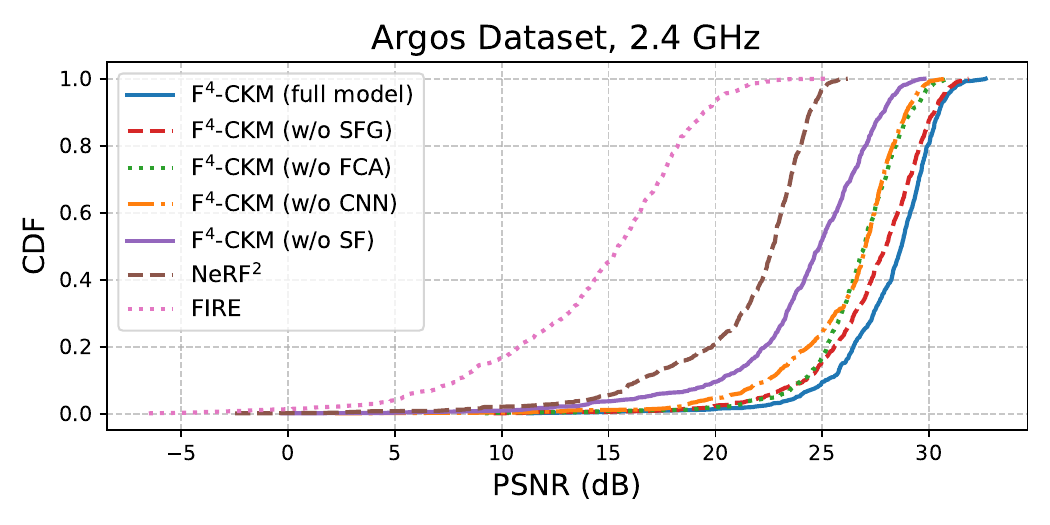}
		\caption{}
		\label{sim_ablation}
	\end{subfigure}
	
	\begin{subfigure}[b]{0.45\textwidth}
		\includegraphics[width=\textwidth]{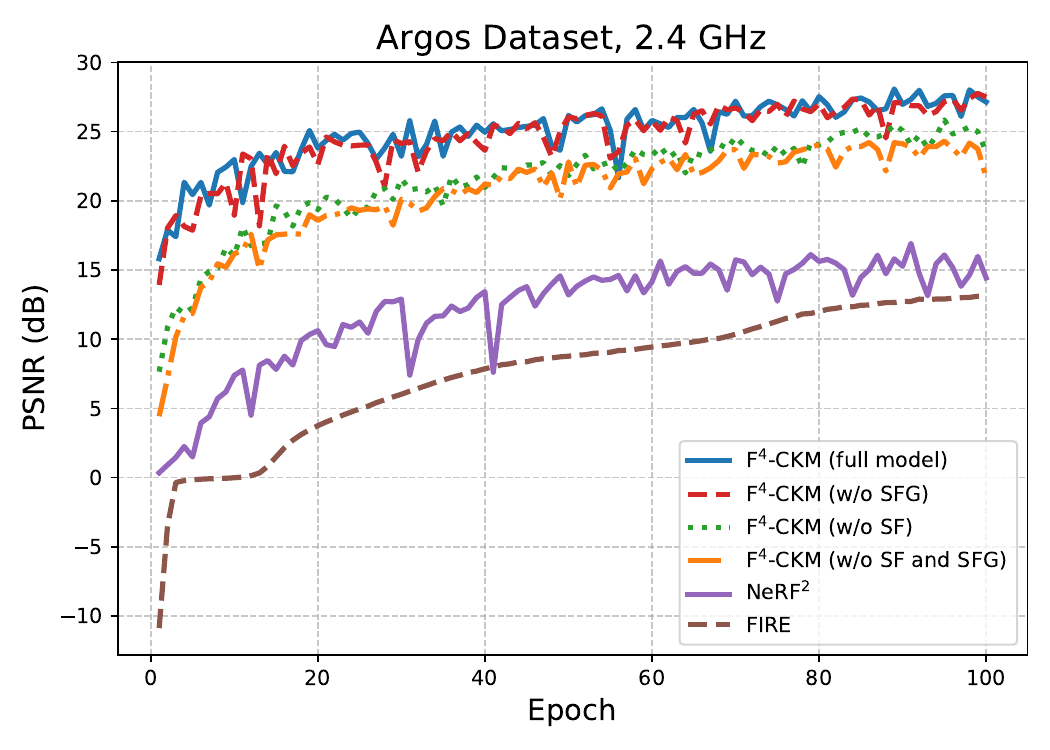}
		\caption{}
		\label{sim_fastlearning}
	\end{subfigure}
	
	\captionsetup{font=footnotesize}
	\caption{Ablation analysis on (a) the PSNR performance and (b) the training efficiency of F$^4$-CKM.}
	\label{sim_ablation_combined}
\end{figure}

\begin{figure}[t]
	\centering
	\includegraphics[width=0.49\textwidth]{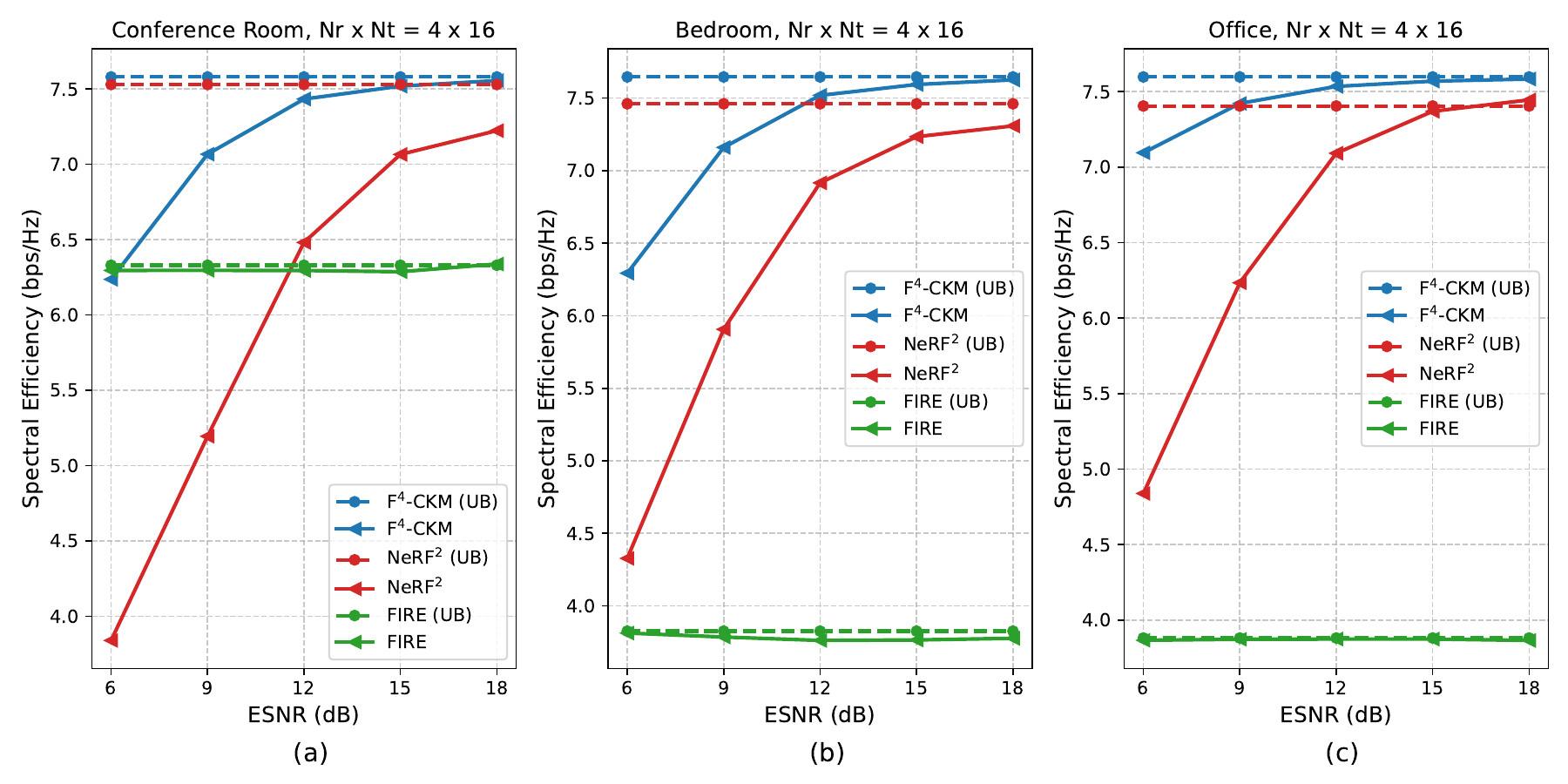}
	\captionsetup{font=footnotesize}
	\caption{The spectral efficiency performance versus the ESNR across three indoor environments. All models are trained with perfect uplink CSI.}
	\label{sim_esnr}
\end{figure}

\begin{figure}[t]
	\centering
	\includegraphics[width=0.45\textwidth]{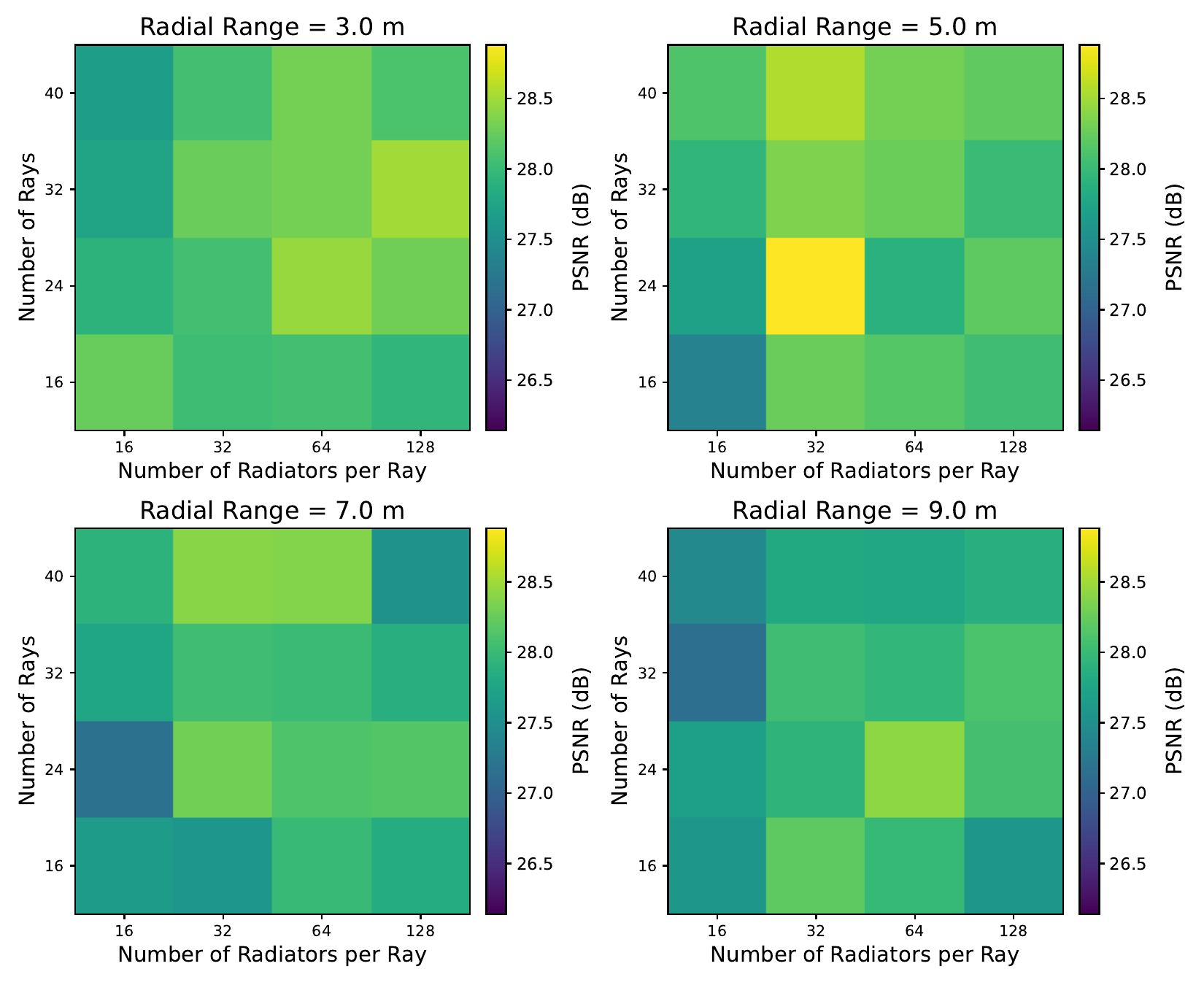}
	\captionsetup{font=footnotesize}
	\caption{The PSNR performance versus the sampling resolution and radial range over the Argos dataset. Each model was trained for $100$ epochs, and the performance was evaluated on the full test set.}
	\label{sim_sampling_res}
\end{figure}

\begin{figure*}[t]
	\centering
	\includegraphics[width=0.85\linewidth]{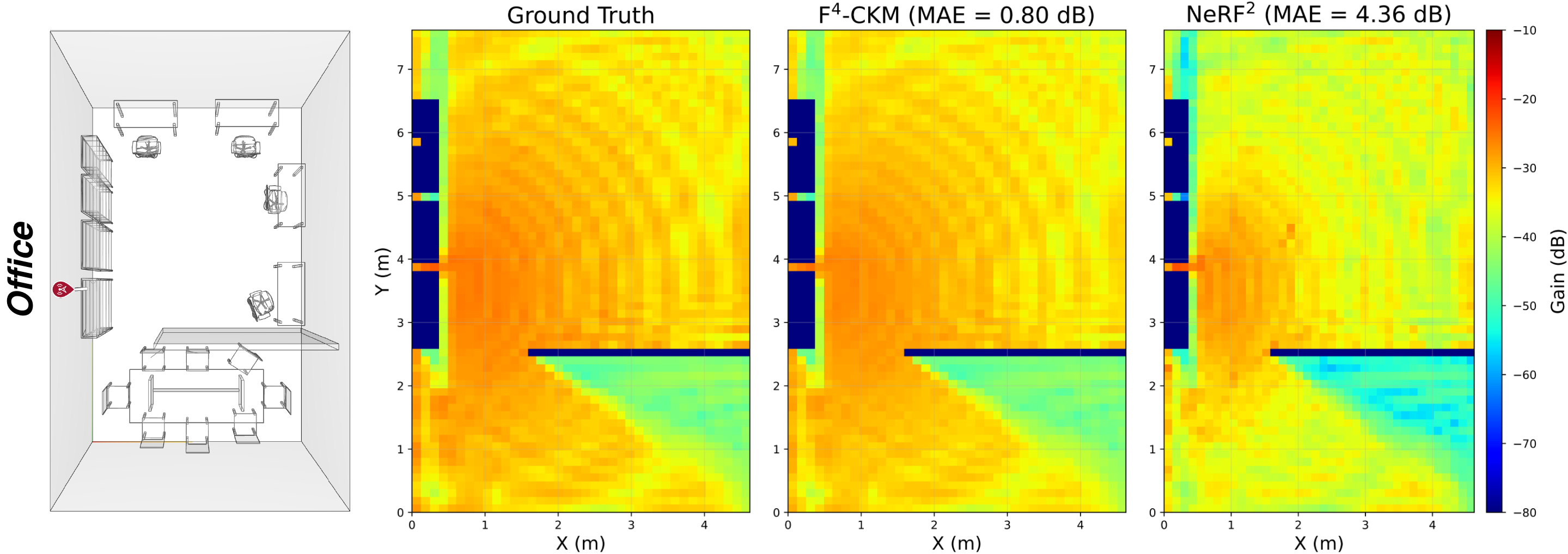}
	\captionsetup{font=footnotesize}
	\caption{Visualization of the learned CKMs. All heatmaps are sampled on a 2D grid with a spacing of one wavelength ($\approx 12.5$ cm at $2.4$ GHz).}
	\label{sim_cgm}
\end{figure*}

\subsection{Ablation Study}

{To systematically evaluate the individual performance contribution of each component in F$^4$-CKM, we conduct a comprehensive ablation study.
As shown in Fig. \ref{sim_ablation_combined}(a), the full F$^4$-CKM model achieves the highest PSNR performance across all percentiles, significantly outperforming all baselines and ablated variants.
Among the ablations, removing the shaping filter (SF) leads to the most severe performance degradation, confirming its critical role in F$^4$-CKM for preserving high-fidelity channel prediction.
Removing the 2D CNN backbone (replaced by an MLP of comparable capacity) incurs the second-largest penalty, underscoring the importance of spatial-aware architectures for accurate prediction.
Disabling the FCA modules results in a moderate loss, indicating that while FCA enhances cross-frequency consistency, capturing spatial context is more essential.
Finally, replacing the SFG sampling strategy with latitude-longitude sampling incurs the smallest performance penalty, suggesting that SFG primarily improves angular uniformity and training efficiency, rather than altering the model’s core representational capability.

Furthermore, Fig. \ref{sim_ablation_combined}(b) evaluates the convergence acceleration provided by the SF and SFG, the only components in F$^4$-CKM explicitly designed for fast learning.
Notably, removing SF severely slows down convergence and degrades the final PSNR, whereas the variant without SFG converges slightly slower but still achieves comparable final performance.
This indicates that SF is the primary driver of fast learning, while SFG acts as an efficiency enhancer that accelerates training without compromising accuracy.
For a comprehensive comparison, the training curves of two baseline methods are also included.
It is worth noting that both the learning rate and batch size keep consistent across all methods.
Notably, F$^4$-CKM boosts an impressive performance leap in the first epoch, nearly surpassing the converged performance of the baselines.
These observations highlight the significant construction efficiency of F$^4$-CKM and indicate its capability for fast fine-tuning and adaptation to new data.}

\subsection{Evaluation with Imperfect Uplink CSI}

In Fig. \ref{sim_esnr}, we examine the {system-level performance} of F$^4$-CKM under imperfect uplink CSI inputs.
To simulate these imperfections, we introduce various levels of additive white Gaussian noise into the uplink CSI inputs.
Additionally, the uplink CSI estimation error is quantified by a metric defined as the estimation signal-to-noise ratio (ESNR), given as
\begin{IEEEeqnarray}{rCl}
	\mathrm{ESNR} = 10\log_{10}(\frac{||\mathbf{H}^\mathrm{U}||^2}{||\mathbf{\hat{H}}^\mathrm{U}-\mathbf{H}^\mathrm{U}||^2}),
\end{IEEEeqnarray}
where $\mathbf{H}^\mathrm{U}$ denotes the actual uplink CSI and $\hat{\mathbf{H}}^\mathrm{U}$ represents the estimated uplink CSI.
{We then evaluate the spectral efficiency performance of F$^4$-CKM versus the ESNR in three indoor environments and compare it with NeRF$^2$ and FIRE.}
All methods are trained with perfect uplink CSI and evaluated across ESNR values ranging from $6$ dB to $18$ dB.
For reference, the performance of all evaluated methods under perfect uplink CSI conditions is also depicted in Fig. \ref{sim_esnr}, labeled as ``upper bound (UB)''.
It is clear that F$^4$-CKM obtains robust performance against uplink CSI estimation errors, highlighting its reliability for practical applications.

{
\subsection{Sensitivity Analysis on Hyperparameters}

In Fig. \ref{sim_sampling_res}, we conduct a systematic sensitivity study across three key hyperparameters, specifically the number of sampled rays, the number of radiators per ray, and the radial sampling range.
Specifically, we trained and evaluated distinct models on the Argos dataset, with each model representing a unique combination of:
\begin{itemize}
	\item \textbf{Number of Rays:} $\{16, 24, 32, 40\}$;
	\item \textbf{Number of Radiators per Ray:} $\{16, 32, 64, 128\}$;
	\item \textbf{Radial Sampling Range:} $\{3.0, 5.0, 7.0, 9.0\}$ m.
\end{itemize}
The results demonstrate that F$^4$-CKM exhibits robust performance across sampling configurations, with the highest value observed at moderate sampling resolution ($24$ rays, $32$ radiators) and with a radial range of $5.0$ m, which we recommend as a strong initial choice for similar scenarios.
For new environments, adaptation can be efficiently achieved via a lightweight grid search over sampling range and resolution.
}

{
\subsection{Visualization of CKM}

In Fig. \ref{sim_cgm}, we provide qualitative visualizations to substantiate the physical grounding of F$^4$-CKM.
Specifically, we generated channel gain heatmaps at $1.2$ m height in the office scenario, comparing the ground-truth channel gain with predictions from F$^4$-CKM and NeRF$^2$.
For reference, we provide the top-down floor plan of the scene, aligned with the heatmap coordinate system, thereby facilitating direct spatial interpretation.
The results demonstrate that F$^4$-CKM captures fine-scale propagation characteristics with high physical fidelity.
Notably, F$^4$-CKM accurately preserves interference patterns near corners and furniture and reproduces shadow regions behind walls, closely matching the ground truth.
The strong geometric consistency observed in F$^4$-CKM's predictions indicates that it learns environment-aware representations governed by wave propagation physics.
}

\section{Conclusion} \label{Conclusion}

In this paper, we introduced F$^4$-CKM, an innovative framework for constructing CKMs with enhanced accuracy and efficiency.
Specifically, we proposed a radiance field rendering-based approach.
First, we formulated a MIMO-OFDM RF radiance field model to capture the spatial-frequency characteristics of wireless channels.
We then developed a novel WiRARE network for RF radiance field rendering, featuring a spatial-aware backbone and enhanced frequency-awareness through FCA modules.
To accelerate convergence, we designed a novel shaping filter module that performs data augmentation on uplink CSI inputs.
This module also enables location-free CKM usage, rendering it resilient to positioning errors.
Additionally, we employed an SFG-based angular sampling strategy to improve efficiency.
Notably, F$^4$-CKM achieves SOTA performance in downlink CSI prediction tasks, significantly outperforming existing methods.
It also exhibits remarkable robustness and high construction efficiency across diverse conditions and environments, highlighting its potential for practical applications.

{
In addition, this paper mainly focuses on indoor static scenarios and extending F$^4$-CKM to large-scale outdoor dynamic environments remains an open challenge.
A key difficulty lies in the substantial increase in required radiator density over kilometer-scale regions, which imposes prohibitive memory and computational costs.
One promising mitigation strategy is an ensemble-and-compression framework: dividing the coverage area into tens-of-meter subregions, deploying a lightweight instantiation of F$^4$-CKM per subregion, and applying model compression (e.g., pruning or quantization) to control storage and inference overhead.
Moreover, in dynamic settings, moving scatters can perturb the radiance distribution, leading to parameter drift and ambiguity in the learned field, thereby degrading prediction fidelity.
To counter this effect, occasional sparse downlink pilot feedback can be used for online recalibration, while its scheduling design and calibration mechanism require further investigation.}

\bibliographystyle{IEEEtran}
\bibliography{IEEEabrv,Reference}

@STRING{TVT         	  = "{IEEE} Trans. Veh. Technol."}

@STRING{JSAC       		  = "{IEEE} J. Sel. Areas Commun."}

@STRING{TCOM      		  = "{IEEE} Trans. Commun."}

@STRING{TWC    		      = "{IEEE} Trans. Wireless Commun."}

@STRING{TCCN      		  = "{IEEE} Trans. Cognit. Commun. Netw."}

@STRING{TMLCN      		  = "{IEEE} Trans. Mach. Learn. Commun. Netw."}

@string{ICC 			   = "Proc. IEEE Int. Conf. Commun. (ICC)"}

@string{WCNC 			   = "Proc. IEEE Wireless Commun. Netw. Conf. (WCNC)"}

@string{VTC 			   = "Proc. IEEE Veh. Technol. Conf. (VTC)"}

@string{PIMRC 			   = "Proc. IEEE Ann. Int. Symp. Pers., Indoor, Mobile Radio Commun. (PIMRC)"}

@string{ECCV 			   = "Proc. Eur. Conf. Comput. Vis. (ECCV)"}

@string{MobiCom 		   = "Proc. ACM Int. Conf. Mob. Comput. Netw. (MobiCom)"}

@string{ICLR 			   = "Proc. Int. Conf. Learn. Represent. (ICLR)"}

@string{ICML 			   = "Proc. Int. Conf. Mach. Learn. (ICML)"}

@ARTICLE{Wu2024transformer,
	author={Wu, Haotian and Shao, Yulin and Ozfatura, Emre and Mikolajczyk, Krystian and Gündüz, Deniz},
	journal=TWC, 
	title={{Transformer}-Aided Wireless Image Transmission With Channel Feedback}, 
	year={2024},
	volume={23},
	number={9},
	pages={11904-11919},
	month={Sep.}}

@ARTICLE{Zhang2024scan,
	author={Zhang, Guangyi and Hu, Qiyu and Cai, Yunlong and Yu, Guanding},
	journal=TCCN, 
	title={{SCAN}: Semantic Communication With Adaptive Channel Feedback}, 
	year={2024},
	volume={10},
	number={5},
	pages={1759-1773},
	month={Oct.}}

@ARTICLE{Chen2023viewing,
	author={Chen, Zirui and Zhang, Zhaoyang and Xiao, Zhuoran and Yang, Zhaohui and Wong, Kai-Kit},
	journal=TWC, 
	title={Viewing Channel as Sequence Rather Than Image: A {2-D} {Seq2Seq} Approach for Efficient {MIMO-OFDM CSI} Feedback}, 
	year={2023},
	volume={22},
	number={11},
	pages={7393-7407},
	month={Nov.}}

@ARTICLE{Wang2021a,
	author={Wang, Jun and Wang, Cheng-Xiang and Huang, Jie and Wang, Haiming and Gao, Xiqi},
	journal=JSAC, 
	title={A General {3D} Space-Time-Frequency Non-Stationary {THz} Channel Model for {6G} Ultra-Massive {MIMO} Wireless Communication Systems}, 
	year={2021},
	volume={39},
	number={6},
	pages={1576-1589},
	month={Jun.}}

@ARTICLE{Lu2022communicating,
	author={Lu, Haiquan and Zeng, Yong},
	journal=TWC, 
	title={Communicating With Extremely Large-Scale Array/Surface: Unified Modeling and Performance Analysis}, 
	year={2022},
	volume={21},
	number={6},
	pages={4039-4053},
	month={Jun.}}

@ARTICLE{You2025next,
	author={You, Changsheng and Cai, Yunlong and Liu, Yuanwei and Di Renzo, Marco and Duman, Tolga M. and Yener, Aylin and Lee Swindlehurst, A.},
	journal=JSAC, 
	title={Next Generation Advanced Transceiver Technologies for {6G} and Beyond}, 
	year={2025},
	volume={43},
	number={3},
	pages={582-627},
	month={Mar.}}

@ARTICLE{Wang2024dynamic,
	author={Wang, Bowen and Li, Hongyu and Cheng, Ziyang},
	journal=TVT, 
	title={Dynamic Hybrid Beamforming Design for Dual-Function Radar-Communication Systems}, 
	year={2024},
	volume={73},
	number={2},
	pages={2842-2847},
	month={Feb.}}

@ARTICLE{Zhao2025energy,
	author={Zhao, Zhouxiang and Yang, Zhaohui and Chen, Mingzhe and Zhu, Chen and Xu, Wei and Zhang, Zhaoyang and Huang, Kaibin},
	journal=TWC, 
	title={Energy-Efficient Probabilistic Semantic Communication over Space-Air-Ground Integrated Networks}, 
	year={2025},
	volume={},
	number={},
	pages={},
	month={early access, May 19,},
	note={{DOI}: {10.1109/TWC.2025.3569102}}}

@ARTICLE{Zhou2025feature,
	author={Zhou, Kequan and Zhang, Guangyi and Cai, Yunlong and Hu, Qiyu and Yu, Guanding and Lee Swindlehurst, A.},
	journal=TWC, 
	title={Feature Allocation for Semantic Communication with Space-Time Importance Awareness}, 
	year={2025},
	volume={},
	number={},
	pages={},
	month={early access, May 20,},
	note={{DOI}: {10.1109/TWC.2025.3569320}}}

@ARTICLE{Wang2025a,
	author={Wang, Bowen and Li, Hongyu and Shen, Shanpu and Cheng, Ziyang and Clerckx, Bruno},
	journal=TCOM, 
	title={A Dual-Function Radar-Communication System Empowered by Beyond Diagonal Reconfigurable Intelligent Surface}, 
	year={2025},
	volume={73},
	number={3},
	pages={1501-1516},
	month={Mar.}}

@ARTICLE{Zeng2021toward,
	author={Zeng, Yong and Xu, Xiaoli},
	journal={IEEE Wireless Commun.},
	title={Toward Environment-Aware {6G} Communications via Channel Knowledge Map},
	year={2021},
	volume={28},
	number={3},
	pages={84-91},
	month={Jun.}}

@ARTICLE{Zeng2024a,
	author={Zeng, Yong and Chen, Junting and Xu, Jie and Wu, Di and Xu, Xiaoli and Jin, Shi and Gao, Xiqi and Gesbert, David and Cui, Shuguang and Zhang, Rui},
	journal={IEEE Commun. Surveys Tuts.},
	title={A Tutorial on Environment-Aware Communications via Channel Knowledge Map for {6G}},
	year={2024},
	volume={26},
	number={3},
	pages={1478-1519},
	month={3rd Quart.}}

@ARTICLE{Liu2025channel,
	title={Channel Knowledge Maps for {6G} Wireless Networks: Construction, Applications, and Future Challenges},
	author={Liu, Xingchen and Sun, Shu and Tao, Meixia and Kaushik, Aryan and Yan, Hangsong},
	journal={arXiv preprint arXiv:2505.24151},
	year={2025}}

@INPROCEEDINGS{Xie2024on,
	author={Xie, Weina and Xu, Xiaoli and Dai, Zhuoyin and Zeng, Yong},
	booktitle=VTC, 
	title={On the Construction of Channel Gain Map: Model-Based or Model-Free Approach?}, 
	year={2024},
	volume={},
	number={},
	pages={},
	month={Jun.},
	address={Singapore, Singapore}}

@ARTICLE{Karttunen2017spatially,
	author={Karttunen, Aki and Molisch, Andreas F. and Hur, Sooyoung and Park, Jeongho and Zhang, Charlie Jianzhong},
	journal=TWC, 
	title={Spatially Consistent Street-by-Street Path Loss Model for {28-GHz} Channels in Micro Cell Urban Environments}, 
	year={2017},
	volume={16},
	number={11},
	pages={7538-7550},
	month={Nov.}}

@INPROCEEDINGS{Kanhere2023calibration,
	author={Kanhere, Ojas and Rappaport, Theodore S.},
	booktitle=ICC, 
	title={Calibration of {NYURay}, a {3D mmWave} and {Sub-THz} Ray Tracer Using Indoor, Outdoor, and Factory Channel Measurements}, 
	year={2023},
	volume={},
	number={},
	pages={5054-5059},
	month={May},
	address={Rome, Italy}}

@ARTICLE{Charbonnier2020calibration,
	author={Charbonnier, Romain and Lai, Chiehping and Tenoux, Thierry and Caudill, Derek and Gougeon, Grégory and Senic, Jelena and Gentile, Camillo and Corre, Yoann and Chuang, Jack and Golmie, Nada},
	journal=TVT, 
	title={Calibration of Ray-Tracing With Diffuse Scattering Against {28-GHz} Directional Urban Channel Measurements}, 
	year={2020},
	volume={69},
	number={12},
	pages={14264-14276},
	month={Dec.}}

@INPROCEEDINGS{Li2022channel,
	author={Li, Kun and Li, Peiming and Zeng, Yong and Xu, Jie},
	booktitle=WCNC, 
	title={Channel Knowledge Map for Environment-Aware Communications: {EM} Algorithm for Map Construction}, 
	year={2022},
	volume={},
	number={},
	pages={1659-1664},
	month={Apr.},
	address={Austin, TX, USA}}

@INPROCEEDINGS{Liu2021fire,
	title={{FIRE}: enabling reciprocity for {FDD MIMO} systems},
	author={Liu, Zikun and Singh, Gagandeep and Xu, Chenren and Vasisht, Deepak},
	booktitle=MobiCom,
	pages={628--641},
	year={2021},
	month={Oct.},
	address={New Orleans Louisiana}}

@ARTICLE{Fu2024generative,
	title={Generative {CKM} Construction using Partially Observed Data with Diffusion Model},
	author={Fu, Shen and Wu, Zijian and Wu, Di and Zeng, Yong},
	journal={arXiv preprint arXiv:2412.14812},
	year={2024}}

@ARTICLE{Si2025unsupervised,
	author={Si, Haonan and Hou, Xiangwang and Wang, Jingjing and Boateng, Gordon Owusu and Zhang, Zekai and Guo, Xiansheng and Niyato, Dusit},
	journal=TWC, 
	title={Unsupervised Localization Toward Crowdsourced Trajectory Data: A Deep Reinforcement Learning Approach}, 
	year={2025},
	volume={},
	number={},
	pages={},
	month={early access, Apr. 30,},
	note={{DOI}: {10.1109/TWC.2025.3563766}}}

@ARTICLE{Jin2025an,
	author={Jin, Zhenzhou and You, Li and Wang, Jue and Xia, Xiang-Gen and Gao, Xiqi},
	journal=TWC, 
	title={An {I2I} Inpainting Approach for Efficient Channel Knowledge Map Construction}, 
	year={2025},
	volume={24},
	number={2},
	pages={1415-1429},
	month={Feb.}}

@ARTICLE{Sifaou2025semi,
	author={Sifaou, Houssem and Simeone, Osvaldo},
	journal=TMLCN,
	title={Semi-Supervised Learning via Cross-Prediction-Powered Inference for Wireless Systems},
	year={2025},
	volume={3},
	number={},
	pages={30-44},
	month={}}

@ARTICLE{Hehn2024differentiable,
	title={Differentiable and Learnable Wireless Simulation with Geometric {Transformers}},
	author={Hehn, Thomas and Peschl, Markus and Orekondy, Tribhuvanesh and Behboodi, Arash and Brehmer, Johann},
	journal={arXiv preprint arXiv:2406.14995},
	year={2024}}

@INPROCEEDINGS{Orekondy2023winert,
	title={{WiNeRT}: Towards neural ray tracing for wireless channel modelling and differentiable simulations},
	author={Orekondy, Tribhuvanesh and Kumar, Pratik and Kadambi, Shreya and Ye, Hao and Soriaga, Joseph and Behboodi, Arash},
	booktitle={Proc. Int. Conf. Learn. Represent. (ICLR)},
	pages={},
	year={2023},
	month={May},
	address={Kigali, Rwanda}}

@ARTICLE{Jin2024sandwich,
	title={{SANDWICH}: Towards an Offline, Differentiable, Fully-Trainable Wireless Neural Ray-Tracing Surrogate},
	author={Jin, Yifei and Maatouk, Ali and Girdzijauskas, Sarunas and Xu, Shugong and Tassiulas, Leandros and Ying, Rex},
	journal={arXiv preprint arXiv:2411.08767},
	year={2024}}

@ARTICLE{Chen2024diffraction,
	author={Chen, Wangqian and Chen, Junting},
	journal=TWC, 
	title={Diffraction and Scattering Aware Radio Map and Environment Reconstruction Using Geometry Model-Assisted Deep Learning}, 
	year={2024},
	volume={23},
	number={12},
	pages={19804-19819},
	month={Dec.}}

@ARTICLE{Hoydis2024learning,
	author={Hoydis, Jakob and Aoudia, Fayçal Aït and Cammerer, Sebastian and Euchner, Florian and Nimier-David, Merlin and Brink, Stephan Ten and Keller, Alexander},
	journal=TMLCN, 
	title={Learning Radio Environments by Differentiable Ray Tracing}, 
	year={2024},
	volume={2},
	number={},
	pages={1527-1539},
	month={}}

@ARTICLE{Xiao2024from,
	author={Xiao, Zhuoran and Zhang, Zhaoyang and Chen, Zirui and Yang, Zhaohui and Huang, Chongwen and Chen, Xiaoming},
	journal=TWC, 
	title={From Data-Driven Learning to Physics-Inspired Inferring: A Novel Mobile {MIMO} Channel Prediction Scheme Based on Neural {ODE}}, 
	year={2024},
	volume={23},
	number={7},
	pages={7186-7199},
	month={Jul.}}

@ARTICLE{Wu2024embracing,
	author={Wu, Guanlin and Lyu, Zhonghao and Zhang, Juyong and Xu, Jie},
	journal={IEEE Open J. Commun. Soc.}, 
	title={Embracing Radiance Field Rendering in {6G}: Over-the-Air Training and Inference With {3-D} Contents}, 
	year={2024},
	volume={5},
	number={},
	pages={4275-4292},
	month={}}

@ARTICLE{Mildenhall2021nerf,
	title={{NeRF}: Representing scenes as neural radiance fields for view synthesis},
	author={Mildenhall, Ben and Srinivasan, Pratul P and Tancik, Matthew and Barron, Jonathan T and Ramamoorthi, Ravi and Ng, Ren},
	journal={Commun. ACM},
	volume={65},
	number={1},
	pages={99--106},
	year={2021},
	month={Dec.}}

@ARTICLE{Kerbl20233d,
	title={{3D Gaussian} splatting for real-time radiance field rendering.},
	author={Kerbl, Bernhard and Kopanas, Georgios and Leimk{\"u}hler, Thomas and Drettakis, George},
	journal={ACM Trans. Graph.},
	volume={42},
	number={4},
	pages={1--14},
	year={2023},
	month={Jul.}}

@INPROCEEDINGS{Zhao2023nerf2,
	title={{NeRF2}: Neural radio-frequency radiance fields},
	author={Zhao, Xiaopeng and An, Zhenlin and Pan, Qingrui and Yang, Lei},
	booktitle={Proc. ACM Int. Conf. Mob. Comput. Netw. (MobiCom)},
	pages={1--15},
	month={Oct.},
	address={Madrid Spain},
	year={2023}}

@ARTICLE{Zhang2024rf,
	title={{RF-3DGS}: Wireless Channel Modeling with Radio Radiance Field and {3D} {Gaussian} Splatting},
	author={Zhang, Lihao and Sun, Haijian and Berweger, Samuel and Gentile, Camillo and Hu, Rose Qingyang},
	journal={arXiv preprint arXiv:2411.19420},
	year={2024}}

@INPROCEEDINGS{Lu2024newrf,
	title={{NeWRF}: A Deep Learning Framework for Wireless Radiation Field Reconstruction and Channel Prediction},
	author={Haofan Lu and Christopher Vattheuer and Baharan Mirzasoleiman and Omid Abari},
	booktitle={Proc. Int. Conf. Mach. Learn. (ICML)},
	pages={33147--33159},
	year={2024},
	month={Jul.},
	address={Vienna, Austria}}

@ARTICLE{Wen2024wrf,
	title={Neural Representation for Wireless Radiation Field Reconstruction: A {3D} {Gaussian} Splatting Approach},
	author={Wen, Chaozheng and Tong, Jingwen and Hu, Yingdong and Lin, Zehong and Zhang, Jun},
	journal={arXiv preprint arXiv:2412.04832},
	year={2024}}

@INPROCEEDINGS{Lee2016comparison,
	author={Lee, Leong and Jones, Matthew and Ridenour, Gregory S. and Bennett, Sam J. and Majors, Arisha C. and Melito, Bianca L. and Wilson, Michael J.},
	booktitle={Proc. IEEE Int. Conf. Comput. Inf. Technol. (CIT)}, 
	title={Comparison of Accuracy and Precision of {GPS}-Enabled Mobile Devices}, 
	year={2016},
	volume={},
	number={},
	pages={73-82},
	month={Dec.},
	address={Nadi, Fiji}}

@ARTICLE{Yang2023environment,
	author={Yang, Yuwen and Gao, Feifei and Tao, Xiaoming and Liu, Guangyi and Pan, Chengkang},
	journal=JSAC, 
	title={Environment Semantics Aided Wireless Communications: A Case Study of {mmWave} Beam Prediction and Blockage Prediction}, 
	year={2023},
	volume={41},
	number={7},
	pages={2025-2040},
	month={Jul.}}

@INPROCEEDINGS{Vieira2017deep,
	author={Vieira, Joao and Leitinger, Erik and Sarajlic, Muris and Li, Xuhong and Tufvesson, Fredrik},
	booktitle=PIMRC, 
	title={Deep convolutional neural networks for massive {MIMO} fingerprint-based positioning}, 
	year={2017},
	volume={},
	number={},
	pages={},
	month={Oct.},
	address={Montreal, QC, Canada}}

@INPROCEEDINGS{Xie2019md,
	title={{mD-Track}: Leveraging multi-dimensionality for passive indoor {Wi-Fi} tracking},
	author={Xie, Yaxiong and Xiong, Jie and Li, Mo and Jamieson, Kyle},
	booktitle=MobiCom,
	pages={1--16},
	year={2019},
	month={Oct.},
	address={Los Cabos Mexico}}

@ARTICLE{Swinbank2006fibonacci,
	title={{Fibonacci} grids: A novel approach to global modelling},
	author={Swinbank, Richard and James Purser, R},
	journal={Quart. J. Roy. Meteorological Soc.},
	volume={132},
	number={619},
	pages={1769--1793},
	year={2006},
	month={Jul.}}

@INPROCEEDINGS{Wu2018group,
	title={Group normalization},
	author={Wu, Yuxin and He, Kaiming},
	booktitle=ECCV,
	pages={3--19},
	year={2018},
	month={Sep.},
	address={Munich, Germany}}

@ARTICLE{Hendrycks2016gaussian,
	title={{Gaussian} error linear units ({GELUs})},
	author={Hendrycks, Dan and Gimpel, Kevin},
	journal={arXiv preprint arXiv:1606.08415},
	year={2016}}

@INPROCEEDINGS{Shepard2016understanding,
	author={Shepard, Clayton and Ding, Jian and Guerra, Ryan E. and Zhong, Lin},
	booktitle={Proc. Asilomar Conf. Signals, Syst. Comput.}, 
	title={Understanding real many-antenna {MU-MIMO} channels}, 
	year={2016},
	volume={},
	number={},
	pages={461-467},
	month={Nov.},
	address={Pacific Grove, CA, USA}}

@ARTICLE{Kingma2014adam,
	title={{Adam}: A method for stochastic optimization},
	author={Kingma, Diederik P and Ba, Jimmy},
	journal={arXiv preprint arXiv:1412.6980},
	year={2014}}

@INPROCEEDINGS{Vasisht2016eliminating,
	title={Eliminating channel feedback in next-generation cellular networks},
	author={Vasisht, Deepak and Kumar, Swarun and Rahul, Hariharan and Katabi, Dina},
	booktitle={Proc. ACM Spec. Interest Group Data Commun. Conf. (SIGCOMM)},
	pages={398--411},
	year={2016},
	month={Aug.},
	address={Florianopolis Brazil}}

@INPROCEEDINGS{Bakshi2019fast,
	title={Fast and efficient cross band channel prediction using machine learning},
	author={Bakshi, Arjun and Mao, Yifan and Srinivasan, Kannan and Parthasarathy, Srinivasan},
	booktitle=MobiCom,
	pages={1--16},
	year={2019},
	month={Oct.},
	address={Los Cabos Mexico}}

@TECHREPORT{3gpp_ts38331,
	author       = {{3GPP}},
	title        = {{NR; Radio Resource Control (RRC); Protocol specification}},
	number       = {38.331},
	type		 = {Technical Specification (TS)},
	institution  = {3rd Generation Partnership Project (3GPP)},
	year         = {2025},
	month		 = {Mar.},
	note         = {Version 18.5.0},
	url          = {https://portal.3gpp.org/desktopmodules/Specifications/SpecificationDetails.aspx?specificationId=3197}
}

\end{document}